\documentclass{aa}  

\usepackage{graphicx}
\usepackage[breaklinks,colorlinks,citecolor=blue]{hyperref}

\usepackage{ulem}
\usepackage{lmodern}

\bibpunct{(}{)}{;}{a}{}{,} 

\begin{document} 

\newif\iffigs

\figstrue

\newcommand{\itii}[1]{{#1}}
\newcommand{\franta}[1]{\textbf{\color{green} #1}}
\newcommand{\itiitext}[1]{{#1}}

\newcommand{\eq}[1]{eq. (\ref{#1})}
\newcommand{\eqp}[1]{(eq. \ref{#1})}
\newcommand{\eqb}[2]{eq. (\ref{#1}) and eq. (\ref{#2})}
\newcommand{\eqc}[3]{eq. (\ref{#1}), eq. (\ref{#2}) and eq. (\ref{#3})}
\newcommand{\refs}[1]{Sect. \ref{#1}}
\newcommand{\reff}[1]{Fig. \ref{#1}}
\newcommand{\reft}[1]{Table \ref{#1}}

\newcommand{\datum} [1] { \noindent \\#1: \\}
\newcommand{\pol}[1]{\vspace{2mm} \noindent \\ \textbf{#1} \\}
\newcommand{\code}[1]{\texttt{#1}}
\newcommand{\figpan}[1]{{\sc {#1}}}

\newcommand{\sfe}{\mathrm{SFE}}
\newcommand{\nbdvi}{\textsc{nbody6} }
\newcommand{\nbdvid}{\textsc{nbody6}}
\newcommand{\mum}{$\; \mu \mathrm{m} \;$}
\newcommand{\rop}{$\rho$ Oph }
\newcommand{\HT}{$\mathrm{H}_2$}
\newcommand{\Halpha}{$\mathrm{H}\alpha \;$}
\newcommand{\HI}{H {\sc i} }
\newcommand{\HII}{H {\sc ii} }
\renewcommand{\deg}{$^\circ$}

\newcommand{\dd}{\mathrm{d}}
\newcommand{\acosh}{\mathrm{acosh}}
\newcommand{\sign}{\mathrm{sign}}
\newcommand{\cex}{\mathbf{e}_{x}}
\newcommand{\cey}{\mathbf{e}_{y}}
\newcommand{\cez}{\mathbf{e}_{z}}
\newcommand{\cer}{\mathbf{e}_{r}}
\newcommand{\ceR}{\mathbf{e}_{R}}

\newcommand{\llg}[1]{\log_{10}#1}
\newcommand{\pder}[2]{\frac{\partial #1}{\partial #2}}
\newcommand{\pderrow}[2]{\partial #1/\partial #2}
\newcommand{\nder}[2]{\frac{\dd #1}{\dd #2}}
\newcommand{\nderrow}[2]{{\dd #1}/{\dd #2}}

\newcommand{\Cmiii}{\, \mathrm{cm}^{-3}}
\newcommand{\Gcmii}{\, \mathrm{g.cm}^{-2}}
\newcommand{\Gcmiii}{\, \mathrm{g.cm}^{-3}}
\newcommand{\Kms}{\, \mathrm{km} \, \, \mathrm{s}^{-1}}
\newcommand{\Si}{\, \mathrm{s}^{-1}}
\newcommand{\Esi}{\, \mathrm{erg} \, \, \mathrm{s}^{-1}}
\newcommand{\Ee}{\, \mathrm{erg}}
\newcommand{\Yr}{\, \mathrm{yr}}
\newcommand{\Myr}{\, \mathrm{Myr}}
\newcommand{\Gyr}{\, \mathrm{Gyr}}
\newcommand{\Msun}{\, \mathrm{M}_{\odot}}
\newcommand{\Rsun}{\, \mathrm{R}_{\odot}}
\newcommand{\Pc}{\, \mathrm{pc}}
\newcommand{\Kpc}{\, \mathrm{kpc}}
\newcommand{\Sd}{\Msun \, \Pc^{-2}}
\newcommand{\Ev}{\, \mathrm{eV}}
\newcommand{\Kk}{\, \mathrm{K}}
\newcommand{\Au}{\, \mathrm{AU}}
\newcommand{\Mas}{\, \mu \mathrm{as}}

   \title{Tidal tails of open star clusters as probes to early gas expulsion II: Predictions for Gaia}


   \author{Franti\v{s}ek Dinnbier
          \inst{1} \&
          Pavel Kroupa \inst{2,3}
          }

   \institute{I.Physikalisches Institut, Universit\"{a}t zu K\"{o}ln, Z\"{u}lpicher Strasse 77, D-50937 K\"{o}ln, Germany \\
             \email{dinnbier@ph1.uni-koeln.de}
         \and
             Helmholtz-Institut f\"{u}r Strahlen- und Kernphysik, University of Bonn, Nussallee 14-16, 53115 Bonn, Germany \\
             \email{pavel@astro.uni-bonn.de}
         \and
             Charles University in Prague, Faculty of Mathematics and Physics, Astronomical Institute, V Hole\v{s}ovi\v{c}k\'{a}ch 2, 180 00 Praha 8, Czech Republic
             }

   \titlerunning{Tidal tails as probes to gas expulsion}
   \authorrunning{F. Dinnbier \& P. Kroupa}

   \date{Received August 25, 2019; accepted February 24, 2020}

 
  \abstract
   {}  
   {We study the formation and evolution of the tidal tail released from 
    a young star Pleiades-like cluster, due to expulsion of primordial gas in a realistic gravitational field of the Galaxy. 
    The tidal tails (as well as clusters) are integrated from their embedded phase for $300 \Myr$.
    We vary star formation efficiencies (SFEs) from $33$\% to $100$\% and the timescales of gas expulsion 
    as free parameters, and provide predictions for the morphology and kinematics of the evolved tail for each of the models. 
    The resulting tail properties are intended for comparison with anticipated 
    Gaia observations in order to constrain the 
    poorly understood early conditions during the gas phase and gas expulsion.}
   {The simulations are performed with the code \nbdvi including a realistic external gravitational 
    potential of the Galaxy, and an analytical approximation for the natal gaseous potential.}
   {Assuming that the Pleiades formed with rapid gas expulsion and an SFE of $\approx 30$ \%, the 
    current Pleiades are surrounded by a rich tail extending from $\approx 150$ to $\approx 350 \Pc$ from the 
    cluster and containing $0.7 \times$ to $2.7 \times$ the number of stars in the present-day cluster. 
    If the Pleiades formed with an SFE close to 100\%, then the tail is shorter ($\lesssim 90 \Pc$) and substantially poorer 
    with only $\approx 0.02 \times$ the number of present-day cluster stars. 
    If the Pleiades formed with an SFE of $\approx 30$ \%, but the gas expulsion was adiabatic, the tail signatures are 
    indistinguishable from the case of the model with 100\% SFE. 
    The mass function of the tail stars is close to that of the canonical mass function for the clusters including primordial gas, 
    but it is slightly depleted of stars more massive than $\approx 1 \Msun$ for the cluster with 100\% SFE, 
    a difference that is not likely to be observed. 
    The model takes into account the estimated contamination due to the field stars and the Hyades-Pleiades stream, 
    which constitutes a more limiting factor than the accuracy of the Gaia measurements.}
   {}


   \maketitle


%

\section{Introduction}

\label{sIntro}


The formation of star clusters commences in infrared dark clouds, which are the densest parts of giant molecular clouds. 
Young massive stars formed within the cloud impart a large amount of energy to the cloud in 
several different forms: ionising radiation \citep{Spitzer1978,TenorioTagle1979,Whitworth1979}, 
stellar winds \citep{Castor1975,Weaver1977}, radiation pressure \citep{Pellegrini2007,Krumholz2009}, 
and later (after $\gtrsim 3 \Myr$) supernovae \citep{Sedov1945a,Taylor1950,McKee1977,Ostriker1988,Cioffi1988,Blondin1998}. 
The plethora of feedback processes opposes and eventually overcomes the self-gravity of the cloud 
while its gaseous content decreases as it turns into stars, 
transforming the initially embedded cluster to a gas free cluster. 
The fraction of gas that transforms to stars is called the star formation efficiency (SFE). 
We use the standard definition of the SFE, $\sfe \equiv M_{\rm cl}/M_{\rm gas}(0)$, where $M_{\rm cl}$ is the total
stellar mass formed from a cloud of initial mass $M_{\rm gas}(0)$ present in the star-forming volume (typically $\approx 1 \Pc^3$).
 We note that this definition of SFE admits a large uncertainty because it is difficult to decide which part of
the cloud is star forming and thus contributes to the mass $M_{\rm gas}(0)$. 

The value of the SFE in current star-forming regions is not well established, 
neither theoretically nor observationally. 
From the theoretical point of view, this is due to the multitude of physical processes included and the very high  spatial 
and temporal resolution of hydrodynamic simulations needed to access this question. 
Consequently, current hydrodynamic simulations resort to various approximations, which results in very different SFEs ranging from $\lesssim 10$\% \citep{Colin2013}, 
$20 - 50$\% \citep{Gavagnin2017,Haid2019}, to models where the early feedback (i.e. before first supernova) is unable 
to disperse a larger fraction of clouds leading to gas exhaustion and SFEs of almost 100\% \citep{Dale2011,Dale2012}. 
As the gas is expelled from the cluster, its gravitational potential lowers, which unbinds a fraction of the stars in the cluster 
or even dissolves the cluster entirely if the SFE is low ($\lesssim 30$\%) and/or the timescale of gas expulsion is very short relative to 
the cluster dynamical time \citep[e.g.][]{Lada1984,Kroupa2001b,Geyer2001}. 
Thus, comparing N-body models where star clusters are impacted by a time-dependent gravitational potential of the gas 
with observed dynamical state of star clusters opens another way for estimating the SFE and the gas expulsion timescale. 
For example, \citet{Banerjee2017} suggest that the large observed size of star clusters at an age of $\approx 2-40 \Myr$ 
cannot be obtained by purely stellar dynamics, but it can be a consequence of rapid gas expulsion and SFEs of $\approx 30$\% (see also \citealt{Pfalzner2009}).

Observationally, only a lower limit of the SFE can be directly obtained from stellar counts because star-forming regions 
are likely to produce more stars from now on before they terminate star formation. 
To directly perform stellar counting is  possible only for the  nearest star-forming regions  of modest mass (typically  several hundred 
solar masses). 
By stellar counting, \citet{Megeath2016} estimate the SFE to be of the order of $10-30$\%, which is consistent with the earlier results of \citet{Lada2003}. 
The expansion of the majority ($\gtrsim 75$\%) of young open star clusters (age $\lesssim 5 \Myr$), as found by \citet{Kuhn2019}, 
indicates a recent violent change of gravitational potential, which likely points to rapid gas expulsion with a low SFE. 
Since this is the second paper in the series, a more extensive introduction to this topic is presented in Dinnbier \& Kroupa (submitted; hereafter Paper I). 

While previous theoretical works concerning star formation focused solely on the star cluster and its related star-forming gas, 
we investigate, for the first time in this paper and in Paper I, 
the properties of the tidal tail composed of stars released during gas expulsion.
Our simulations also self-consistently include the younger tidal tail which arises from stars released
due to dynamical processes (evaporation and ejections) later in the exposed cluster.

Since models of different SFEs and rapidity of gas expulsion 
release a different number of stars and the stars are released at different speeds, 
the predicted tidal tails are substantially different between the models. 
These simulations are intended mainly to give predictions for the tidal tails to be observed around 
open star clusters by the Gaia mission. 
The parameters of our models are tailored to the \object{Pleiades} open star cluster; however, 
they can be easily extrapolated to other and more massive clusters, which is also discussed. 
From the properties of the observed tidal tails (or their absence), it will be 
possible to constrain the appropriate initial conditions, and thus the SFE during the cluster formation.


\section{Numerical methods and initial conditions}

\label{sInitCond}

The present clusters extend the parameter space of the models studied in Paper I. 
Since the setting of the code \nbdvi \citep{Aarseth1999,Aarseth2003} is described in Paper I, 
here we only briefly summarise the cluster initial conditions and the model of gas expulsion. 

The stars in clusters follow the Plummer distribution of total cluster mass $M_{\rm cl}$ and 
the Plummer scale-length $a_{\rm cl}$. 
Stellar masses are distributed according to the initial mass function (IMF) of \citet{Kroupa2001a} in the mass range of ($0.08 \Msun$, $m_{\rm up}$), 
where $m_{\rm up}$ is chosen from the $m_{\rm up} - M_{\rm cl}$ relation of \citealt{Weidner2010,Weidner2013} 
(see also the earlier works of \citealt{Elmegreen1983,Elmegreen2000}), 
which implies $m_{\rm up} = 40 \Msun$ for the $1400 \Msun$ clusters, and $m_{\rm up} = 80 \Msun$ for the $4400 \Msun$ clusters.
The cluster moves around the Galaxy on a circular orbit of radius $8.5 \Kpc$; the Galactic potential 
is taken from \citet{Allen1991}.

The gravitational potential of the natal gas is represented by a Plummer model 
of initial mass $M_{\rm gas}(0)$ and time-independent Plummer scale-length $a_{\rm cl}$. 
After time $t_{\rm d}$, $M_{\rm gas}$ is reduced exponentially on timescale $\tau_{\rm M}$. 
We use $t_{\rm d} = 0.6 \Myr$ for all models. 
For the models with rapid (impulsive) gas expulsion, we adopt $\tau_{\rm M} = a_{\rm cl}/(10 \Kms)$ as in Paper I.
Models with slow (adiabatic) gas expulsion have $\tau_{\rm M} = 1\Myr$.
The parameters of the clusters are listed in \reft{tsimList}.

It should be noted that the values of $t_{\rm d}$ and $\tau_{\rm M}$ are only roughly constrained 
from observations and (to a lesser extent) from
hydrodynamic simulations.
Particularly the gas expulsion timescale $\tau_{\rm M}$ might be
significantly longer or shorter than $a_{\rm cl}/(10 \Kms)$, which is given by the dominant physical mechanism
responsible for the feedback and its interplay with gravity,
which has not been satisfactorily understood.
The dependence of the star cluster dynamical state on the parameter $\tau_{\rm M}$ was  
previously studied by \citet{Baumgardt2007} and \citet{Brinkmann2017}.

\begin{table*}
\begin{tabular}{cccccccccc}
Model name & $M_{\rm cl} (0)$ & $M_{\rm gas} (0)$ & $r_{\rm h}(0)$ & $\tau_{M}$ & $t_{\rm h}$ & $t_{\rm rlx}$ & $t_{\rm ms}$ & $\widetilde{v}_{\rm e,I}$ & $\widetilde{v}_{\rm e,II}$  \\
 & [$\Msun$] & [$\Msun$] & [$\Pc$] & [$\Myr$] & [$\Myr$] & [$\Myr$] & [$\Myr$] & [$\Kms$] & [$\Kms$] \\
\hline
C03G13 & 1400 & 2800 & 0.20 & 0.020 & 0.028 & 13.1 & 0.6 & 2.3 & 1.0 \\
C03G23 & 1400 &  700 & 0.20 & 0.020 & 0.034 & 5.5  & 0.3 & 1.6 & 1.0 \\
C03GA  & 1400 & 2800 & 0.20 & 1.000 & 0.028 & 13.1 & 0.6 & 1.6 & 0.8 \\
C03W1  & 1400 &    0 & 1.00 &   -   & 0.43  & 38   & 1.8 &  -  & 0.8 \\
C03W5  & 1400 &    0 & 5.00 &   -   & 4.8   & 430  & 20  &  -  & 0.6 \\
\textit{C03W02}  & 1400 &    0 & 0.2 &   -   & 0.037 & 3.3  & 0.2 & - & \\
\hline
C10G13 & 4400 & 8800 & 0.23 & 0.020 & 0.020 & 25.2 & 1.2 & 4.0 & 1.4 \\
C10G23 & 4400 & 2200 & 0.23 & 0.020 & 0.023 & 10.6 & 0.5 & 3.2 & 1.4 \\
C10GA  & 4400 & 8800 & 0.23 & 1.000 & 0.020 & 25.2 & 1.2 & 3.8 & 1.3 \\
C10W1  & 4400 &    0 & 1.00 &   -   & 0.24  & 58   & 2.7 &  -  & 1.2 \\
C10W5  & 4400 &    0 & 5.00 &   -   & 2.6   & 650  & 30  &  -  & 1.1 \\
\textit{C10W02}  & 4400 &    0 & 0.23 &   -   & 0.026 & 6.3  & 0.3 & - & \\
\end{tabular}
\caption{List of star cluster models. 
From left to right, the meaning of the columns is as follows: 
 model name, initial stellar mass of the cluster $M_{\rm cl}(t=0)$, initial mass $M_{\rm gas}(t=0)$ of the gas component, 
initial half-mass radius $r_{\rm h}(0)$, gas expulsion timescale $\tau_{M}$, 
half-mass crossing time $t_{\rm h}$, relaxation timescale $t_{\rm rlx}$, and  timescale for mass segregation 
for $10 \Msun$ stars $t_{\rm ms}$. 
The last two columns give the median escape speed of stars in tails I and II 
($\widetilde{v}_{\rm e,I}$ and $\widetilde{v}_{\rm e,II}$), respectively.
The difference in the timescales for clusters with the same stellar mass and radius (e.g. for models 
C03G13 and C03W02) stems from different velocity dispersion due to the gaseous potential.
The models performed  for only one simulation are in italics.}
\label{tsimList}
\end{table*}

\section{Evolution of the star cluster models, escaping stars and the onset of the tidal tail formation}

\label{sOverviewModels}

In this section the main emphasis is on understanding the 
relation between the dynamical impact of the gas expulsion mechanism on cluster dynamics 
and the formation and evolution of the tidal tail.
Here the tidal tail properties are described mainly quantitatively, 
and the qualitative investigation of the tidal tail is described in Sect. \ref{sTail}.

The models start at cluster formation as deeply embedded objects, 
follow gas expulsion of their natal gas, which unbinds a significant fraction of stars, and integrate 
both the star cluster members and the tidal tail for $300 \Myr$, when the clusters are dynamically evolved. 
The stars released during gas expulsion form tidal tail I, the kinematics of which is 
investigated in Paper I. 
The dynamical evolution of the star cluster then continues for the whole simulation, and gradually unbinds other 
stars, due to evaporation and ejections, forming tidal tail II. 

We particularly focus on how the initial conditions of the gas expulsion influence the properties of the tidal tails, 
and the fraction of stars present in each tail and in the star cluster.  
Although the results can be applied to any star cluster, we particularly aim at the Pleiades, due to its proximity, and 
plot the quantities of interest at the age of the Pleiades $t_{\rm pl}$, which we take to be $125 \Myr$ \citep{Stauffer1998}.

The cluster models attempt to capture the representative 
cases of the rich diversity of initial conditions 
admitted by the still limited knowledge of the initial state of young star clusters. 
The models are listed in Table \ref{tsimList}. 
The initial cluster mass $M_{\rm cl}(0)$ is chosen so that the mass of the cluster at time $t_{\rm pl}$ 
is comparable to the current mass of the Pleiades; each scenario for gas expulsion or gas free cluster 
is calculated for a lower mass ($M_{\rm cl}(0) = 1400 \Msun$) and a more massive ($M_{\rm cl}(0) = 4400 \Msun$) cluster. 
We use models with fast (impulsive; $\tau_{\rm M} \ll t_{\rm h}$ where $t_{\rm h}$ is the half-mass crossing time) 
gas expulsion and $\sfe = 1/3$ (models C03G13 and C10G13), fast gas expulsion 
and $\sfe = 2/3$ (C03G23 and C10G23), and with slow (adiabatic; $\tau_{M} \gg t_{\rm h}$) gas expulsion and $\sfe = 1/3$ (C03GA and C10GA). 
The initial half-mass radii $r_{\rm h}$ are adopted from the relation between the embedded cluster 
mass and its radius, $r_{\rm h} = 0.1 \Pc \; (M_{c}/\Msun)^{1/8}$, which is suggested by \citet{Marks2012}, 
which gives $r_{\rm h} = 0.2$ and $r_{\rm h} = 0.23 \Pc$ for the lower mass and more massive cluster, respectively.
We also calculate models without gas, but with initial half-mass radii $r_{\rm h}$ chosen so that they bracket 
the half-mass radii of the models C03G13 and C10G13 after gas expulsion and revirialisation.
The gas-free models have initially $r_{\rm h} = 1 \Pc$ (models C03W1 and C10W1) and 
$r_{\rm h} = 5 \Pc$ (C03W5 and C10W5). 
In addition, to discuss the roles of stellar dynamics and gas expulsion on the early cluster expansion, 
we study gas-free clusters with the same initial half-mass radius as the gaseous models (models C03W02 and C10W02). 

In order to obtain better statistics, we repeat each model of the more massive cluster four times, and
lower mass cluster 13 times with a different random seed 
\footnote{Models C03W02 and C10W02 are performed only for one realisation, and they are not analysed 
to the same extent as the other models.}
(to have a comparable number of stars for the analysis
in the lower mass and more massive clusters, i.e. $4\times4400 \Msun \approx 13 \times 1400 \Msun$). 
Unless stated otherwise, the figures and tables below are the average over all the realisations. 

Since lower and higher mass models often show the same response on changing the gas expulsion prescription, 
and the cluster mass plays a subordinate role for the results qualitatively, we often describe the results 
of models with different masses together, and omit the cluster mass from the generic name (e.g. CG13 refers to 
two models, C03G13 and C10G13).

Table \ref{tsimList} also lists relevant dynamical timescales: the half-mass crossing time $t_{\rm cr} = r_{\rm h}/\sigma_{\rm cl}$, 
where $\sigma_{\rm cl}$ is the velocity dispersion of the cluster;
the half-mass relaxation time (see \citealt[][their eq. 7.108]{Binney2008}),
%
\begin{equation}
t_{\rm rlx} = \frac{0.17 N}{\ln(\lambda N)}\sqrt{\frac{r_{\rm h}^3}{G }},
\label{etrlx}
\end{equation}
where $N$ is the number of stars, using the value of the Coulomb logarithm $\lambda = 0.1,$ 
as recommended by \citet{Giersz1994}, and
$t_{\rm ms}$ is the mass segregation timescale $t_{\rm ms}$ for massive stars of mass $m_2$, embedded in a sea of 
low mass stars of mass $m_1$: $t_{\rm ms} = 0.9 (m_1/m_2) t_{\rm rlx}$ (\citealt[][his eq. 9]{Spitzer1969}; 
see also \citealt{Spitzer1940,Mouri2002}). 
We use the mass of the lower mass group of stars to be the mean mass of the stars in the cluster, $m_1 = 0.47 \Msun$, 
while the massive stars have $m_2 = 10 \Msun$. 

When analysing the kinematic properties of the star cluster and the tidal tail, 
we consider only the objects for which good statistics can be obtained, so we 
discard compact objects. 
The analysis takes into account red giants and main sequence stars. 
When referring to all stars, we count each star as one object regardless of its mass; 
when referring to groups of stars (according to their spectral class), we likewise 
do not take into account differences of stellar masses within the group and assign each star the same weight. 

\iffigs
\begin{figure*}
\includegraphics[width=0.94\textwidth]{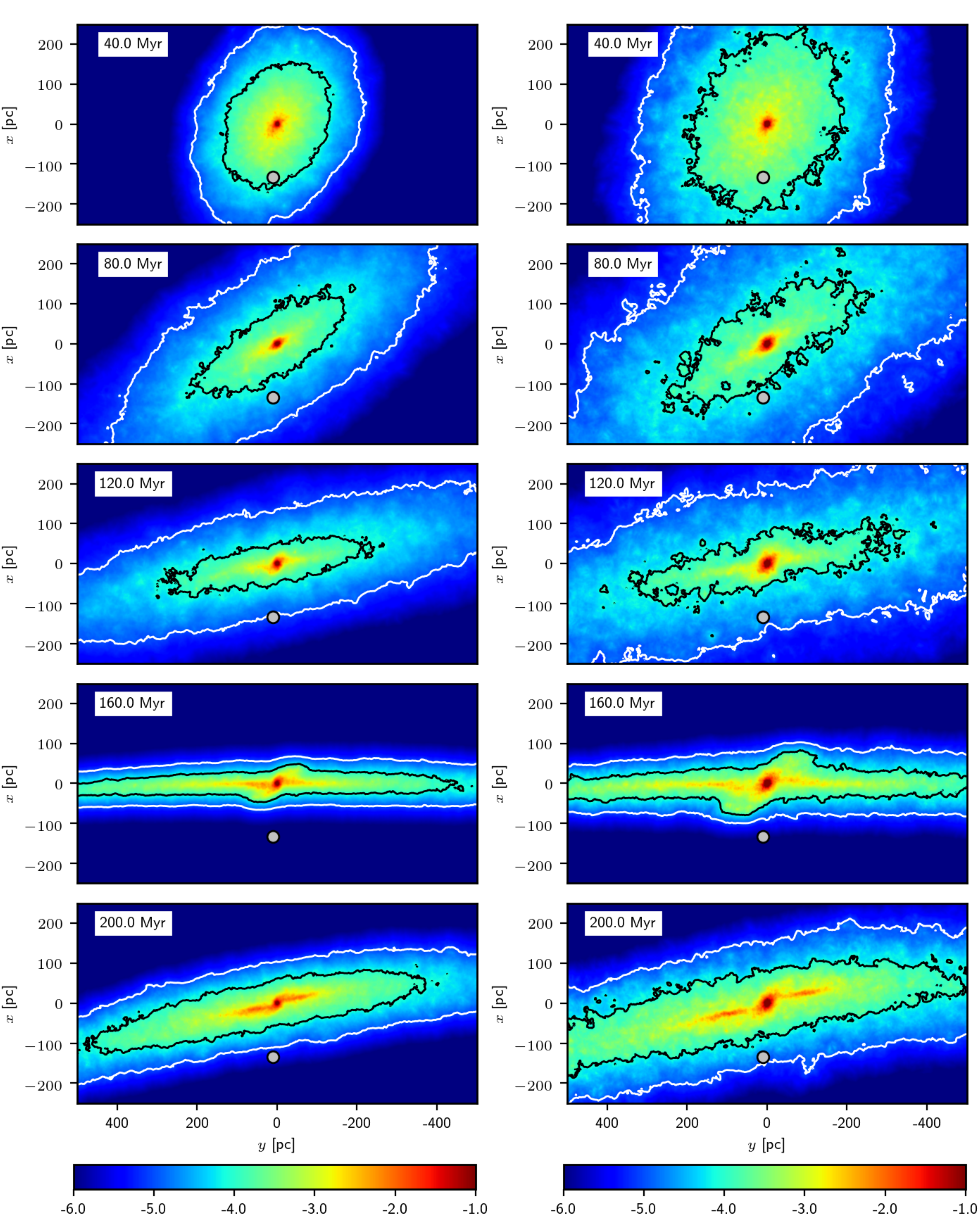}
\caption{Evolution of the star number density in plane $z = 0$ for the models with rapid gas 
expulsion and $\sfe=1/3$ (model C03G13, left; model C10G13,  right). 
The time is indicated in the upper left corner of each frame. 
The positive $y$-axis points in the direction of 
the Galactic rotation, the negative $x$-axis points towards the Galactic centre (the view is from the south Galactic pole). 
The position of the Sun is indicated by the large grey circle. 
The colour scheme shows the number density of stars in units $\llg{(\mathrm{pc}^{-3}).}$
Each plot is an average of between 4 (for the more massive models C10G13) and 13 (for the less massive models C03G13) 
simulations with a different random seed. 
The  white contours   indicate the contamination threshold for the Besan\c{c}on model 
represented by the Schwarzschild velocity distribution ($n_{\rm tl} = 1.0 \times 10^{-5} \Pc^{-3}$) 
and the black contours that for the Besan\c{c}on model and Hyades-Pleiades stream combined ($n_{\rm tl} = 9.0 \times 10^{-5} \Pc^{-3}$; here we take three times 
the value calculated in Sect. \ref{ssContamin} to get an upper estimate of contamination).
}
\label{fevg13}
\end{figure*} \else \fi

\iffigs
\begin{figure*}
\includegraphics[width=\textwidth]{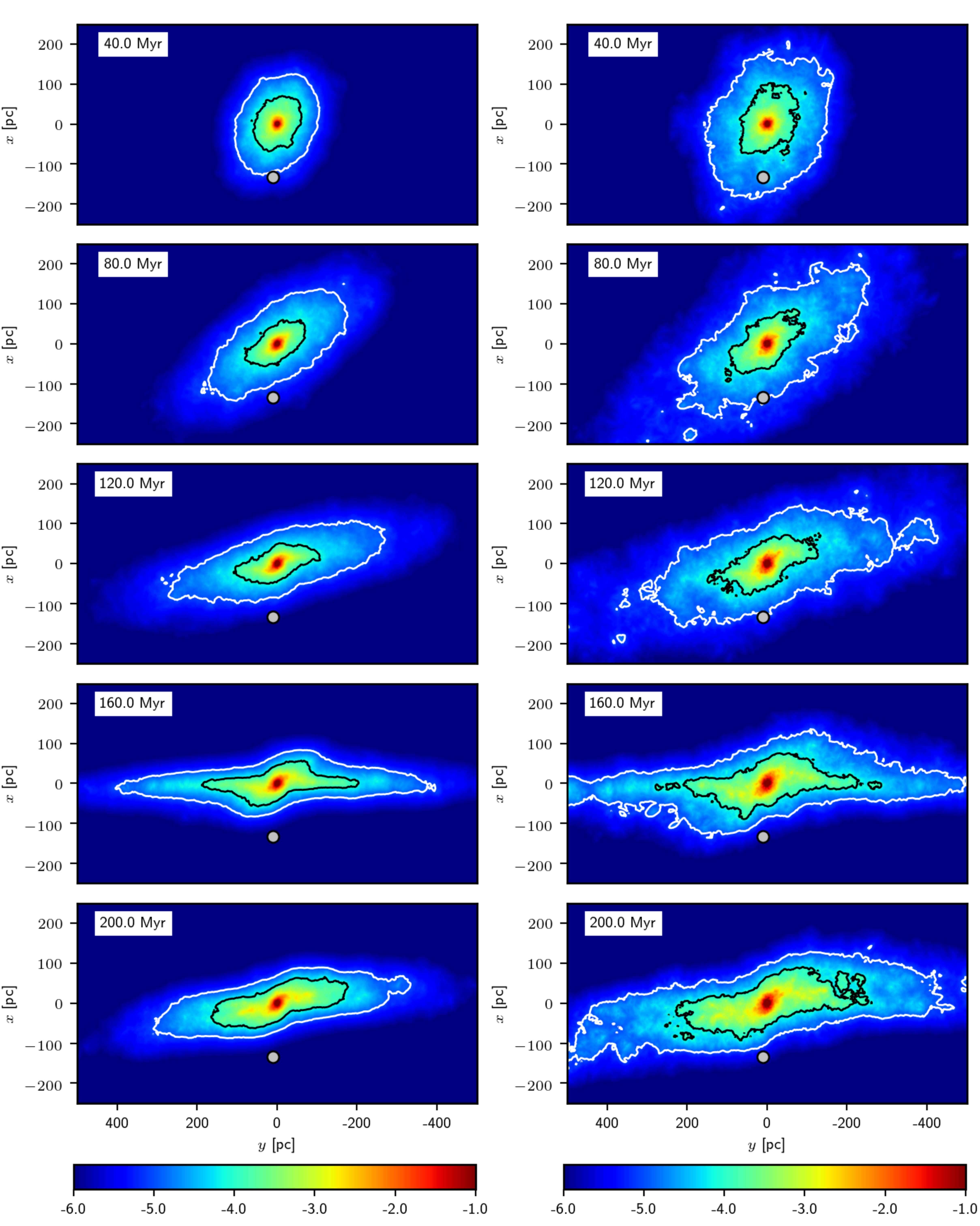}
\caption{Same as Fig. \ref{fevg13}, but for the models with rapid gas expulsion and $\sfe=2/3$ 
(model C03G23,  left; model C10G23, right).}
\label{fevg23}
\end{figure*} \else \fi

\iffigs
\begin{figure*}
\includegraphics[width=\textwidth]{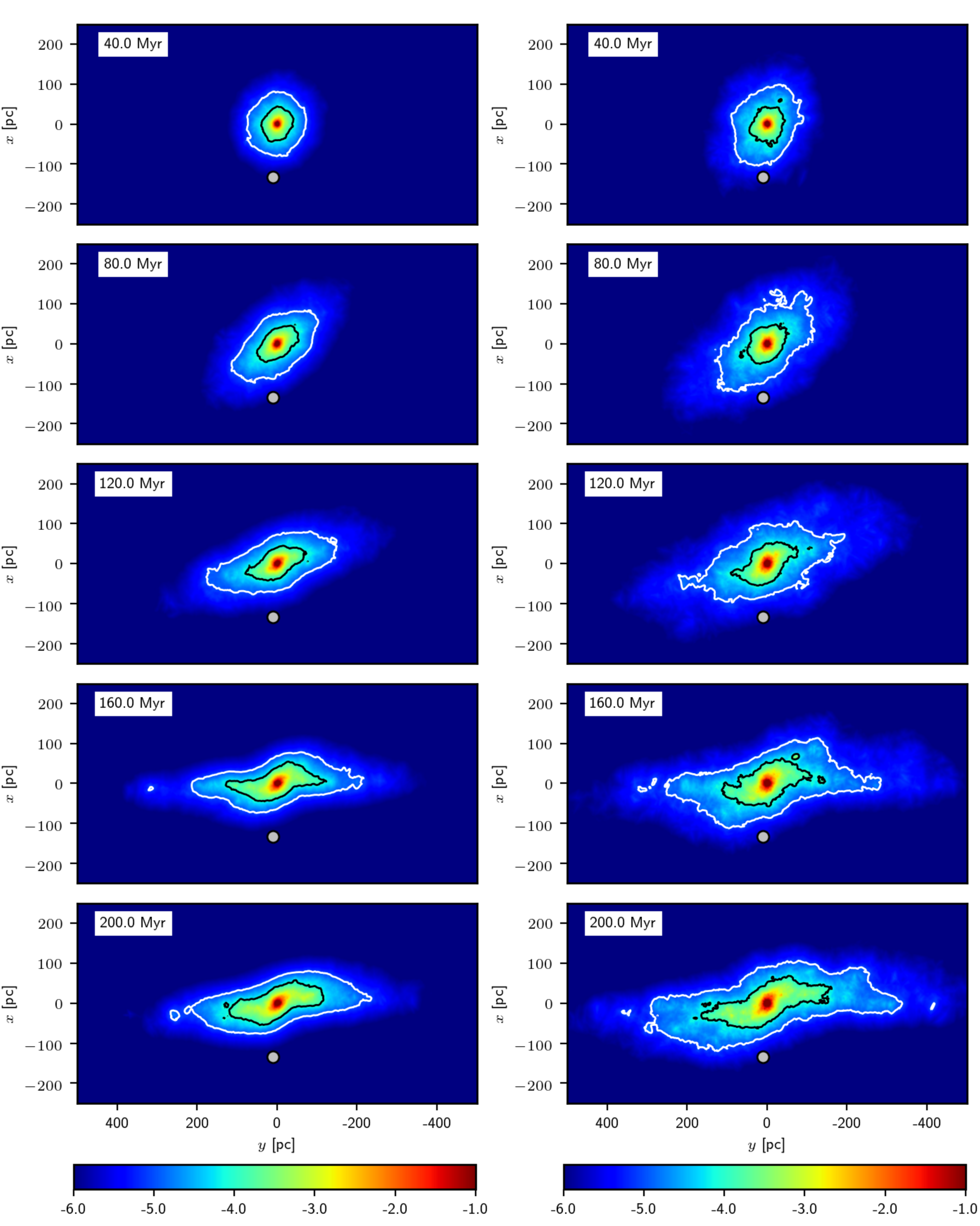}
\caption{Same as Fig. \ref{fevg13}, but for the models with adiabatic gas expulsion and $\sfe=1/3$ 
(model C03GA; the left column, model C10GA; the right column).}
\label{fevAdiab}
\end{figure*} \else \fi

\iffigs
\begin{figure*}
\includegraphics[width=\textwidth]{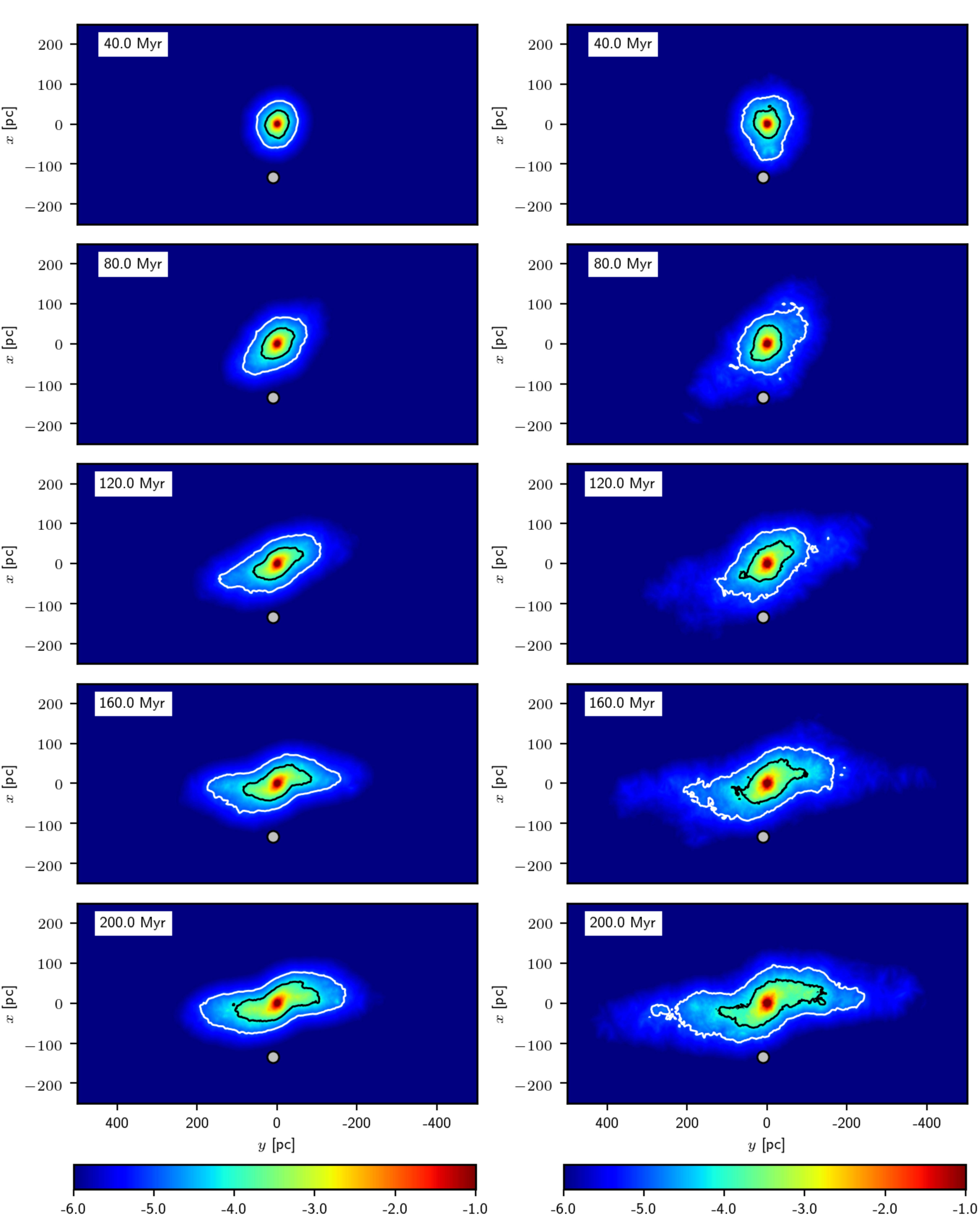}
\caption{The same as Fig. \ref{fevg13} for the models without primordial gas and with $r_{\rm h} (0) = 1 \Pc$ 
(model C03W1; the left column, model C10W1; the right column).}
\label{fevWo1}
\end{figure*} \else \fi

\iffigs
\begin{figure*}
\includegraphics[width=\textwidth]{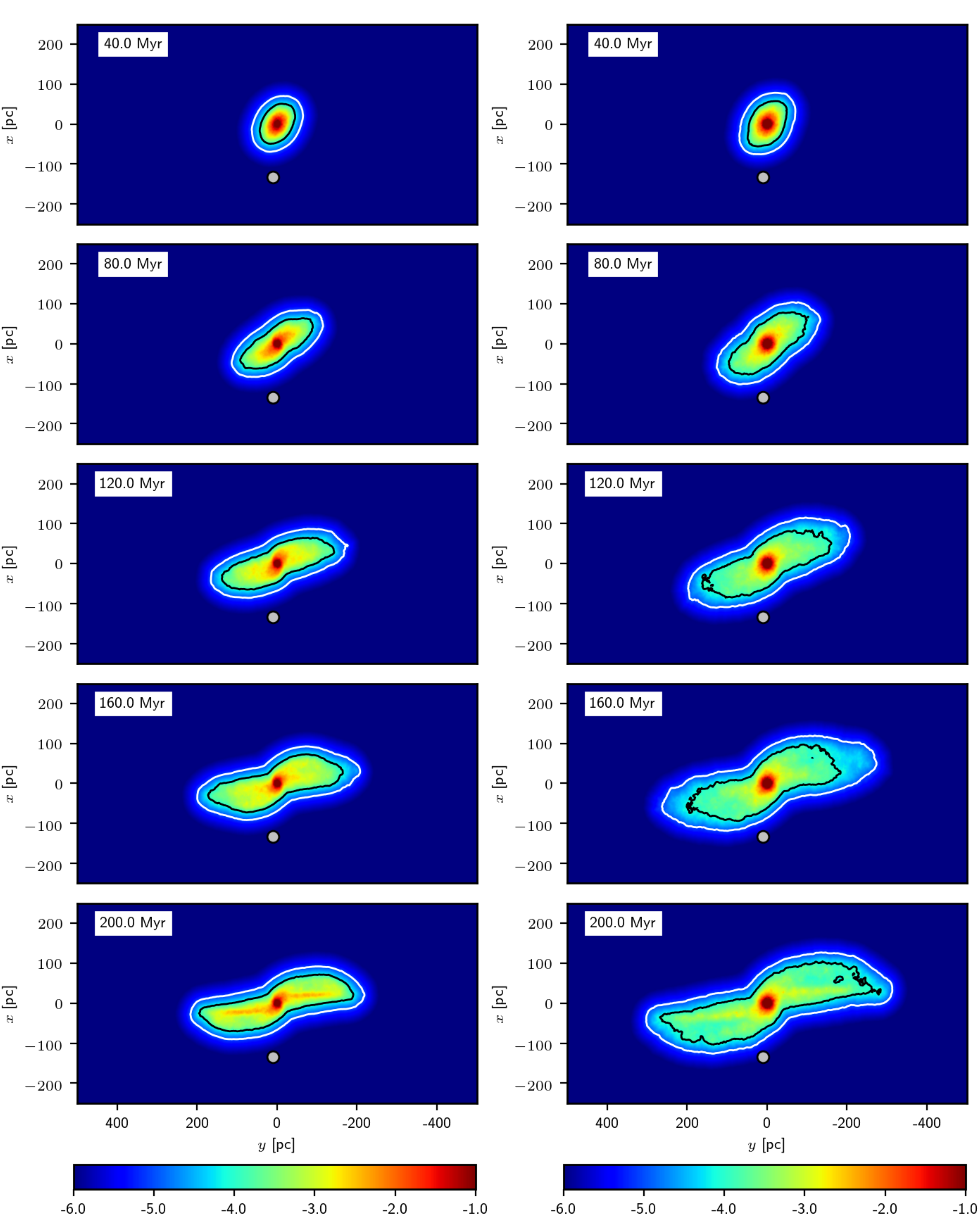}
\caption{The same as Fig. \ref{fevg13} for the models without primordial gas and with $r_{\rm h} (0) = 5 \Pc$
(model C03W5; the left column, model C10W5; the right column).}
\label{fevWo5}
\end{figure*} \else \fi

Figures \ref{fevg13} through \ref{fevWo5} show the evolutionary sequence of the stellar number density at the Galactic 
midplane $z = 0$ for each model at age $40 \Myr$ to $200 \Myr$ in intervals of $40 \Myr$. 
The left column corresponds to the lower mass clusters ($M_{\rm cl}(0) = 1400 \Msun$), while 
the right column corresponds to the more massive clusters ($M_{\rm cl}(0) = 4400 \Msun$).
 The stellar number density, which is shown by the colour scheme, is calculated 
by the method developed by \citet{Casertano1985}, where
the radius of the sphere is set according to the position of the sixth closest star to the point where the density is evaluated.
Although these plots hold for any cluster of comparable mass and similar orbit, 
we indicate the relative position of the Sun by the  grey circle for the particular case of the Pleiades. 
The contours represent the density of field stars, which contaminate the possible observation, for the 
case of the Pleiades for the Schwarzschild velocity distribution (white contour) and the Schwarzschild velocity 
distribution and Hyades-Pleiades stream combined 
(black contour). The field star contamination is discussed in detail in Sect. \ref{ssContamin}.

\iffigs
\begin{figure*}
\includegraphics[width=\textwidth]{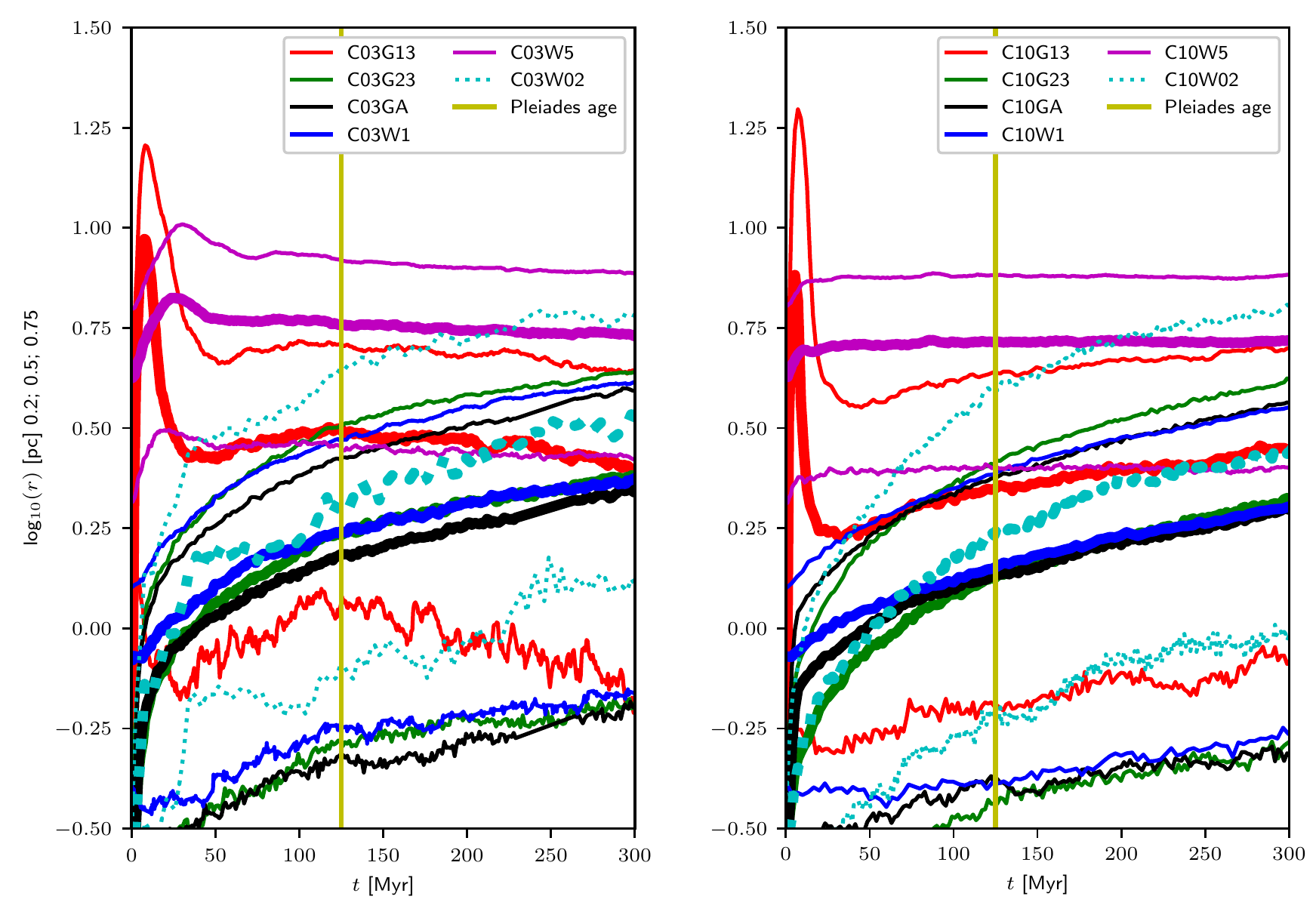}
\caption{Evolution of the 0.2, 0.5 (thick line), and 0.75 Lagrangian radius of the star cluster models.
Shown are the lower ($M_{\rm cl}(0) = 1400 \Msun$) and higher mass ($M_{\rm cl}(0) = 4400 \Msun$) clusters, (left and right panels, respectively).
The age of the Pleiades is indicated by the vertical yellow line.  }
\label{fLagrangeCl}
\end{figure*} \else \fi
 
The membership of a star to tails I and II 
depends on the time when the star escaped the cluster in comparison to the time $t_{\rm ee} = 2 r_{\rm t}/(0.1 \sigma_{\rm cl} (t = 0))$.
Stars escaping before $t_{\rm ee}$ belong to tail I, while stars escaping after $t_{\rm ee}$ belong to  
tail II; this is the same criterion we used in Paper I.
The total number of tail stars, $N_{\rm tl}$,  at the age of the Pleiades, $t_{\rm pl}$, is provided in \reft{tTailAbund}. 
The evolution of the fraction of the tail stars, $N_{\rm tl}$, to all stars in 
the system (i.e. $N_{\rm tot} = N_{\rm tl} + N_{\rm cl}$, where $N_{\rm cl}$ is the number of cluster stars) 
is shown in the upper panel of \reff{fsigmaNStars}. 
The fraction of tail I stars within the tail (i.e. $N_{\rm tl,I}/N_{\rm tl}$) is shown in the middle panel of the figure. 

Before investigating properties of the tails for all the models, 
we recall the main differences between tails I and II, as discussed in Paper I. 
Tail I is formed from faster escaping stars (see the velocity $\widetilde{v}_{\rm e,I}$ in \reft{tsimList}), 
which all escape during a short time interval. 
Tail I expands faster, and its thickness in directions $x$ and $z$ as well as velocity dispersion oscillate aperiodically. 
During the maxima of tail $x$ thickness, tail I forms a thick envelope around the cluster while it collapses 
to a narrow strip near the minimum of $x$ thickness. 
Its expanding motion stops temporarily, and there are time intervals when the tail shrinks towards the cluster. 
In contrast, tail II expands gradually at a slower speed, is thinner most of the time, and is of a distinct S-shape. 
Tail II also forms overdensities around the epicyclic cusps \citep{Kupper2008}.
Thus, the tidal tails of the models containing gas (CG13, CG23 and CGA) are a superposition of tail I and tail II, 
but with different numbers of stars and different stellar escape speeds in each of the tails. 


\subsection{Clusters with SFE 33\% and rapid gas expulsion}

The clusters are strongly impacted by the loss 
of gas, which causes the cluster to expand rapidly. 
The expansion is illustrated on Lagrangian radii in Fig. \ref{fLagrangeCl}. 
The half-mass radius expands by almost two orders of magnitude (from $\approx 0.2 \Pc$ to $\approx 10 \Pc$, and 
then revirialises to $\approx 2-3 \Pc$. 
See also that the revirialisation is faster for more massive clusters, and that the half-mass radius
for more massive clusters is smaller after revirialisation. 
The cluster velocity dispersion, $\sigma_{\rm cl}$, abruptly decreases as the cluster revilialises after the rapid expansion
(lower panel of Fig. \ref{fsigmaNStars}), and the value of $\sigma_{\rm cl}$ is then close to that of  some models without primordial gas.
We note that the behaviour of 
the models with SFE 33\% shows the same typical evolutionary phases as the models studied in the past by 
 \citet[][]{Kroupa2001b}, \citet{Banerjee2013}, and \citet{Brinkmann2017}, among others. 

\iffigs
\begin{figure}
\includegraphics[width=0.99\columnwidth]{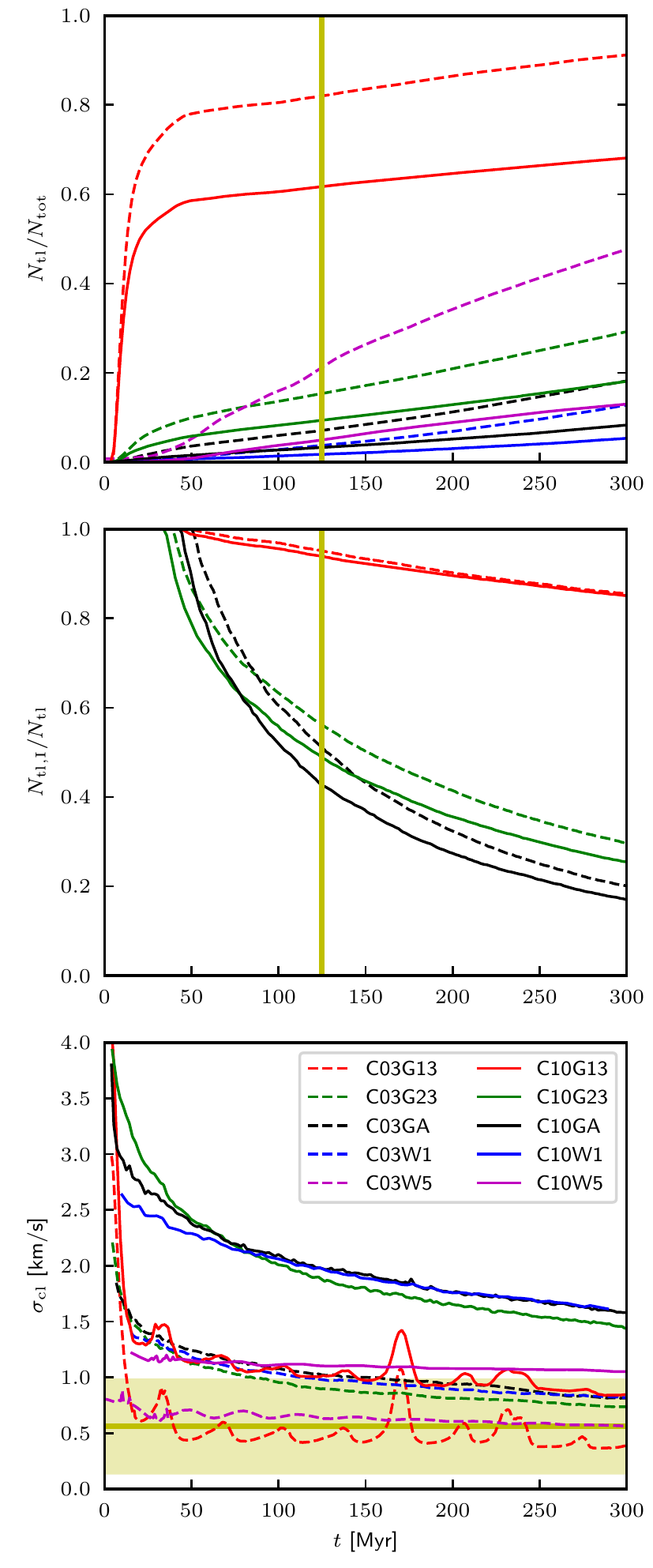}
\caption{\figpan{Upper panel:} Fraction of all stars in the tail, $N_{\rm tl}$, normalised to the total number 
of stars, $N_{\rm tot}$, as a function of time. 
The results for lower mass and more massive clusters are shown by dashed and solid lines, respectively. 
The age of the Pleiades is indicated by the vertical yellow line.
\figpan{Middle panel:} Fraction of the tail stars $N_{\rm tl,I}/N_{\rm tl}$ which belong to tail I. 
\figpan{Lower panel:} Time dependence of the velocity dispersion $\sigma_{\rm cl}$ for stars within the star cluster. 
The observed velocity dispersion for the Pleiades and its $1 \sigma$ error according to \citet{Raboud1998} 
is indicated by the horizontal yellow band. }
\label{fsigmaNStars}
\end{figure} \else \fi

The rapid gas expulsion unbinds around 75 \% (for model C03G13) or 60 \% (for model C10G13) of all stars from the cluster 
(upper panel of Fig. \ref{fsigmaNStars}). 
The following evaporation and occasional ejections are far less effective processes in unbinding stars 
than the gas expulsion. 
This results in the tidal tail to be dominated by tail I (middle panel of \reff{fsigmaNStars}). 
At the age of the Pleiades, tail I comprises $95$ \% of tail stars, and it still comprises $85$ \% of 
tail stars at the age of $300 \Myr$ (middle panel of Fig. \ref{fsigmaNStars}). 
This suggests that the tidal tail evolves closely to tail I, which is studied in Paper I. 

The evolutionary sequence of the tidal tail, as shown in Fig. \ref{fevg13}, 
 demonstrates the aperiodic oscillations in thickness with first minimum at the age of $\approx 160 \Myr$, 
when the tail aligns with axis $y=0$ as follows from eq. 15 in Paper I.
We note that the two models evolve qualitatively similarly;  the more massive cluster, as  expected, forms
a longer and a wider tail because its stars were released at a greater speed.

Tail II becomes prominent later (see the snapshot at $160 \Myr$ in Fig. \ref{fevg13}), 
where tail II is shorter (length $\approx 100 \Pc$) and tilted at an angle $\approx 45 ^\circ$ to the $y$-axis.
Later, tail II transforms to the characteristic S-shape (the frame at $200 \Myr$), 
which is well understood from studies of gas-free clusters \citep[e.g.][]{Kupper2008,Kupper2010}.
Tail II owes its slower growth to the lower velocity of the stars composing it relative to the
stars in tail I (see Table \ref{tsimList}). 
Thus, a star cluster starting with an $\sfe=1/3$ and rapid gas expulsion forms a rich tidal tail with 
an oscillating thickness and tilt to the $y$-axis,  and a less populated and shorter tidal tail due to evaporation, which
is  S-shaped and non-oscillating.

\subsection{Clusters with SFE 66\% and rapid gas expulsion}

The increase in SFE from 1/3 to 2/3 has a dramatic influence on the cluster evolution. 
Gas expulsion plays a far smaller role, unable to  impact the cluster significantly. 
The cluster gradually expands (by factor of $\approx 5$ from the beginning to $t_{\rm pl}$) 
without any sign of revirialisation (green lines in Fig. \ref{fLagrangeCl}).
The cluster velocity dispersion gradually decreases (lower panel of Fig. \ref{fsigmaNStars}). 

In order to estimate the role of gas expulsion in addition to pure stellar dynamics, 
we calculate an extra model with the same initial conditions as these clusters but without any gas component. 
We run  one realisation of this model for a cluster with 
$M_{\rm cl}(0) = 1400 \Msun$ (model C03W02) and one with $M_{\rm cl}(0) = 4400 \Msun$ (model C10W02). 
From Fig. \ref{fLagrangeCl} it follows that these models evolve closely to models with $\sfe = 2/3$ up to $\approx 50 \Myr$ (i.e. far longer than the gas expulsion timescale $\tau_{\rm M}$), indicating that the role of the external gaseous potential 
plays a rather  minor role if $\sfe = 2/3$ is assumed, and that the cluster evolution is dominated by 
pure N-body evolution, where the  expansion of the cluster radius is driven by escapers coming from 
violent interactions in small (several bodies) temporary systems near the cluster centre \citep{Fujii2011,Tanikawa2012}.

The modest influence of the gas expulsion unbinds far fewer stars than in the cluster with $\sfe = 1/3$; 
only around 9 \% to 15 \% (for model C03G23 and C10G23, respectively) of all stars are present in the 
tail by the age of the Pleiades (see the upper panel of Fig. \ref{fsigmaNStars}). 
In contrast to the models CG13, 
both tails I and II are composed of a comparable number of stars (cf. $N_{\rm tl,I}$ and $N_{\rm tl}$ 
in the middle panel of \reff{fsigmaNStars}).

Apart from the less numerous tail I, 
which is also released at a slower speed $\widetilde{v}_{\rm e,I}$ (see Table \ref{tsimList}), 
the evolution of the tidal tail bears distinct similarities to the tail of the model with $\sfe = 1/3$. 
Tail I first expands approximately radially, and then tilts and aligns with the $y$-axis at $\approx 160 \Myr$ 
(the evolutionary sequence for models C03G23 and C10G23 is shown in Fig. \ref{fevg23}). 
Tail II slowly builds up independently of tail I, and gradually assumes an S shape  
(see the panels after $\approx 120 \Myr$ in Fig. \ref{fevg23}).

\subsection{Clusters with SFE 33\% and adiabatic gas expulsion}

The evolution of the  half-mass radius during the first $10 \Myr$ is shown in Fig. \ref{flagrcldetail}. 
For comparison, we plot an estimate of the half-mass radius $r_{\rm h}$ assuming 
conservation of adiabatic invariants, 
i.e. $r_{\rm h}(t) \approx r_{\rm h}(0) (M_{\rm cl}(0) + M_{\rm gas}(0))/(M_{\rm cl}(0) + M_{\rm gas}(t))$ 
(\citealt{Hills1980,Mathieu1983}; dotted yellow line). 
The more massive model C10GA expands rapidly during the first $\approx 4\Myr$ with $r_{\rm h}$ somewhat lower than 
from the adiabatic invariant, but catches up with it at $\approx 6 \Myr$. 
We attribute the smaller expansion of $r_{\rm h}$ relatively to the analytical estimate from collisional dynamics. 
This is even more relevant for the smaller cluster (model C03GA), which has a shorter relaxation time, 
where the expansion of $r_{\rm h}$ is less pronounced than in the more massive cluster (model C10GA). 
We note that the expansion cannot be caused by non-equilibrium initial conditions because in this case 
the cluster would expand on the crossing timescale $t_{\rm h}$, which is $\approx 0.04 \Myr$ (\reft{tsimList}), 
while the expansion occurs on a substantially longer timescale of several Myr, which is 
comparable to the relaxation timescale $t_{\rm rlx}$. 
Model C10GA expands to a larger radius than model C10G23 (after $3 \Myr$), which contains less gas, but after the 
initial expansion both models reach comparable radii around $\approx 50 \Myr$ (Fig. \ref{fLagrangeCl}) and then evolve close to each 
other (see also $\sigma_{\rm cl}$ in the lower plot of Fig. \ref{fsigmaNStars}). 
The slower expansion of model C10GA than C10G23 after $100 \Myr$ is due to a slower dynamical evolution 
of cluster C10GA, which gets diluted early. 

Although tail I contains $\approx 1/2$ of the total tail population 
at the age of the Pleiades (middle panel of Fig. \ref{fsigmaNStars}), 
which is similar to models CG23, the tidal tail is less numerous (the upper panel), containing only $3-5$ \% of all stars.  
The evolutionary snapshots (Fig. \ref{fevAdiab}) indicate that tail I is practically absent, 
and the tail is dominated by tail II slowly forming its S-shape structure. 
Thus, apart from the early inflation of the cluster, adiabatic gas expulsion has very little 
impact on the cluster or tail evolution.

\subsection{Gas-free clusters with $r_{\rm h}(0) = 1 \Pc$}

These  models (C03W1 and C10W1) gradually expand due to their internal N-body dynamics (Fig. \ref{fLagrangeCl}). 
Because of their longer relaxation time (they are only $2t_{\rm rlx}$ to $4 t_{\rm rlx}$ old at the age of the Pleiades; 
see Table \ref{tsimList}), the rate of their initial expansion is smaller 
than that of the models with $\sfe = 2/3$ or the models with adiabatic gas expulsion. 

The models slowly build an S-shaped tidal tail (see Fig. \ref{fevWo1}) that is free of any 
tilting and pulsating component, as we saw in tails for the models with rapid gas expulsion.
From all the models studied, these clusters produce the least populated tails, containing 
only $1\%-2$\% of the cluster population by the Pleiades age (see the upper panel of Fig. \ref{fsigmaNStars}). 

\subsection{Gas-free clusters with $r_{\rm h}(0) = 5 \Pc$}

With respect to the previous clusters, the large initial half-mass radius of $5 \Pc$ means that these models 
(C03W5 and C10W5) almost fill their tidal radii ($r_{\rm t} = 16 \Pc$ and $24 \Pc$, respectively), which results in more rapid evaporation. 
The half-mass radius (as well as the $0.2$ and $0.75$ Lagrangian radii) remain nearly constant (Fig. \ref{fLagrangeCl}) 
as the outwardly moving stars  escape rather than remain in the cluster when reaching a larger distance from the 
cluster centre. 
The cluster velocity dispersion also remains  nearly constant (lower panel of Fig. \ref{fsigmaNStars}), or 
very slowly decreases as the cluster mass decreases (it decreases more for the lower mass model C03W5, which evaporates faster; 
see also \citealt{Baumgardt2003}). 

\iffigs
\begin{figure*}
\includegraphics[width=\textwidth]{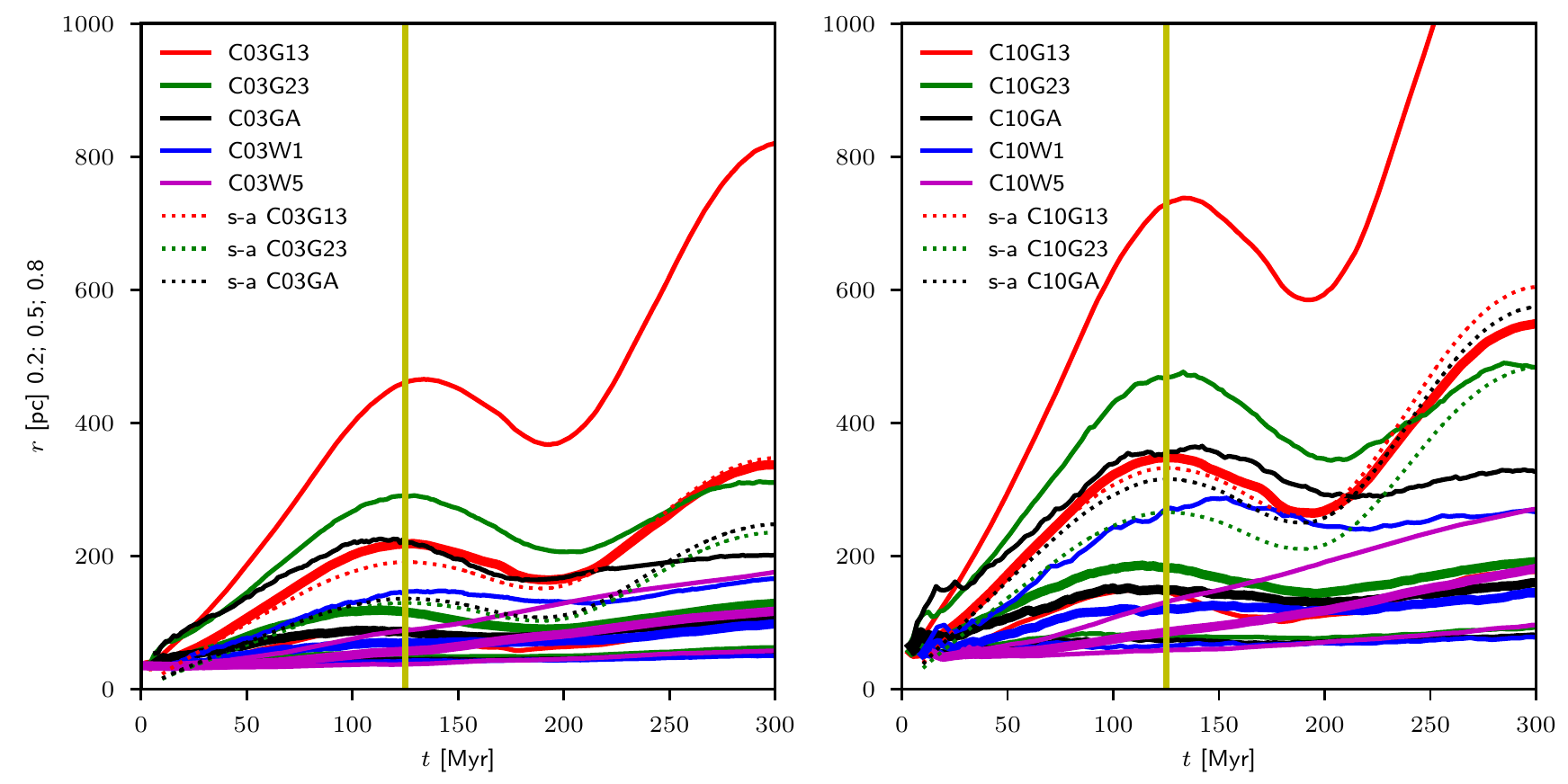}
\caption{Evolution of the 0.2, 0.5, and 0.8 Lagrangian radius of the tidal tail for number of stars (solid lines). 
Left and right panels represent the lower ($M_{\rm cl}(0) = 1400 \Msun$) 
and higher mass ($M_{\rm cl}(0) = 4400 \Msun$) clusters, respectively. 
The half-number radius (i.e. the 0.5 Lagrangian radius) is indicated by the thick solid lines. 
The semi-analytical half-number radius calculated by eq. 20 in Paper I and scaled to the 
value of $\widetilde{v}_{\rm e,I}$ (see \reft{tsimList}) for the corresponding model is shown by the dotted lines. 
The semi-analytical prediction is very good for the models dominated by tail I (models C03G13 and C10G13; red lines), 
but it is rather poor for the models with a non-negligible population of tail II stars (models C03G23, C10G23, C03GA, and C10GA).}
\label{flagrtail}
\end{figure*} \else \fi

The tidal tail contains $20$\% and $5$\% of all cluster stars at the Pleiades age for models C03W5 and C10W5, respectively, being 
richer than the tail of the more concentrated gas-free models CW1. 
The evolution of the tail is shown in Fig. \ref{fevWo5}. 
The tail is clearly S-shaped (tail II), without any other component, so it is distinct from the tails formed 
in the models with $\sfe = 1/3$ and 2/3 and rapid gas expulsion. 
At a given distance from the cluster, the tail is substantially denser than that in models CW1; 
it is so because the tail is not only more populated, but also of a lower velocity relative to the cluster. 
We note that although the isodensity contours of the tail are placed at a larger distance from the cluster than in models 
CW1 (cf. Figs. \ref{fevWo1} and \ref{fevWo5}), the tail is actually shorter in models CW5 
(see the half-number radius $r_{\rm h,tl}$ of the tail in \reff{flagrtail} below). 
Models CW1 have stars at relatively large distances from the clusters, but these stars are not 
shown in Fig. \ref{fevWo1} because they fall below the density threshold.

\section{Evolution of the tidal tails}

\label{sTail}

\subsection{Physical extent and kinematics}

\label{ssTailKine}

\iffigs
\begin{figure*}
\includegraphics[width=\textwidth]{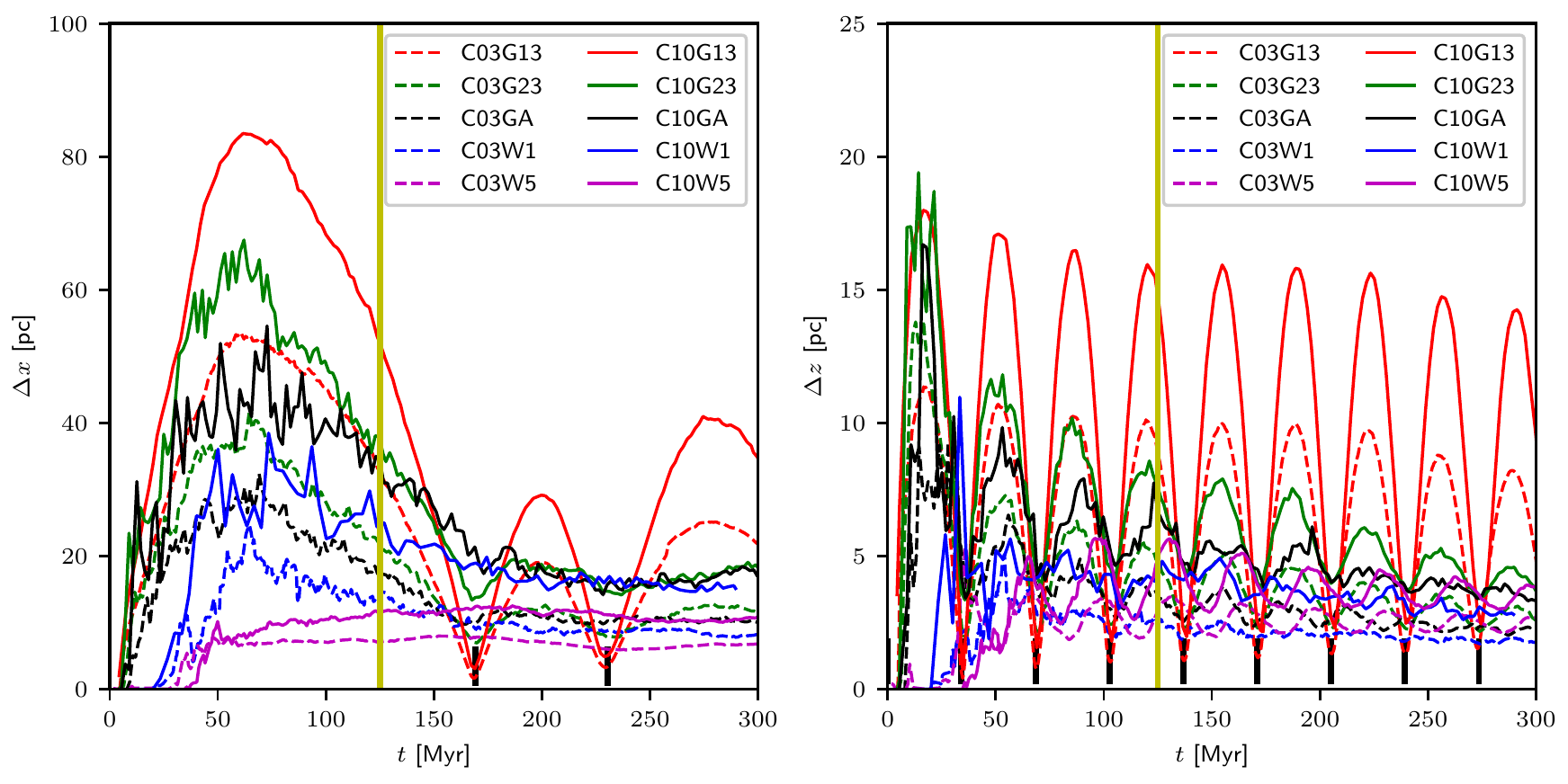}
\caption{Evolution of the tail thickness measured in the strip of $200 \Pc < y < 300 \Pc$ in direction $x$ (left panel) 
and in direction $z$ (right panel).  
The age of the Pleiades is shown by the yellow vertical line. 
The minima of thickness for the analytic model described in Sect. 2.2 of Paper I are
indicated by the short black vertical bars.}
\label{ftailthickness}
\end{figure*} \else \fi

\iffigs
\begin{figure*}
\includegraphics[width=\textwidth]{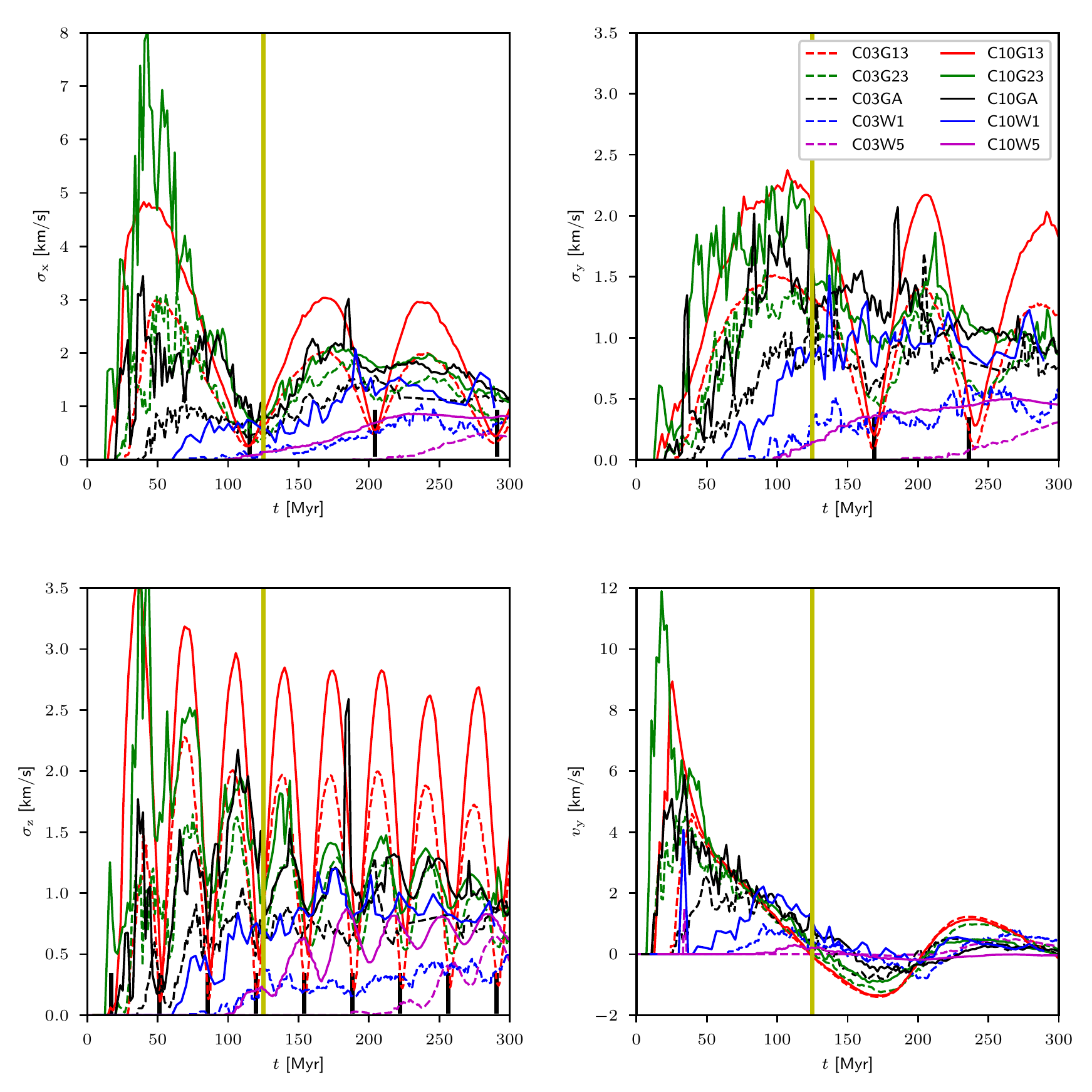}
\caption{Evolution of the velocity structure of the tidal tail at $200 \Pc < y < 300 \Pc$. 
\figpan{Upper left:} Velocity dispersion in  direction $x$. 
\figpan{Upper right:} Velocity dispersion in  direction $y$. 
\figpan{Lower left:} Velocity dispersion in  direction $z$. 
\figpan{Lower right:} Bulk velocity in direction $y$. 
The short black vertical bars indicate the minima for the corresponding quantity according to the 
semi-analytic model of Paper I. 
The yellow vertical lines show the age of the Pleiades.}
\label{fvelvstime}
\end{figure*} \else \fi

Although the present work focuses on the Pleiades, we study the tidal tails for a longer time of $300 \Myr$ also to 
provide predictions for more star clusters. 
This section extends the results of Paper I, where we study two extreme scenarios of gas expulsion (models C10G13 and C10W1),
to scenarios with different SFEs or $\tau_{\rm M}$ and  to lower mass clusters.  

The extent of the tail is approximated by its $50$\% Lagrangian radius. 
The Lagrangian radii are calculated for a given number of stars, not for the mass to avoid fluctuations caused by 
a small statistics of massive stars; it is the same approach as adopted in Paper I. 
The tails are stretched along the $y$-axis due to the Coriolis force, so the Lagrangian radius is close to the tail 
extent along line $y$ (cf. Figs. \ref{fevg13} through \ref{fevWo5}). 
The models with rapid gas expulsion and $\sfe = 1/3$ (CG13) form the most extended tails 
with the half-number radius $r_{\rm h,tl}$ of $220 \Pc$ 
and $350 \Pc$ at the age of the Pleiades (\reff{flagrtail}; see also \reft{tContamin}). 
Models C03G23 and C10G23 have shorter tails at $t_{\rm pl}$ ($r_{\rm h,tl} = 180 \Pc$ and $120 \Pc$, respectively). 
The other models form even shorter tails ($r_{\rm h,tl} \approx 120 \Pc$ and $\approx 70 \Pc$ at $t_{\rm pl}$ for the more and less massive 
model, respectively). 

In Paper I, we find that the semi-analytic estimate for the tail half-number radius 
is in  very good agreement with $r_{\rm h,tl}$ of model C10G13, which is dominated by tail I. 
Here, we study how good this approximation is for the tails of models CG23 and CGA, which 
have a larger fraction of tail II stars. 
The results of eq. 20 in Paper I scaled to the median escape speed $\widetilde{v}_{\rm e,I}$ for tail I stars (\reft{tsimList}) 
are shown by dotted lines in \reff{flagrtail}. 
The semi-analytic model significantly overestimates (by a factor of $\approx 2-3$) the value of $r_{\rm h,tl}$ 
for the clusters with a non-negligible tail II population (i.e. models CG23 and CGA)  because of the smaller distances of tail II stars to the cluster, which have
a smaller escape speed $\widetilde{v}_{\rm e,II}$.

Next we study the evolution of the tail thickness, tail velocity dispersion and the bulk velocity of the tail 
in a narrow strip located at $|y' - y| < \Delta y$. 
The restriction to the strip is in order to focus on the time variability 
of these quantities for given $y$ instead of spatial dependence along the $y$-axis. 
The spatial dependence along the $y$-axis of these quantities for the extreme models 
of gas expulsion scenarios (models C10G13 and C10W1)  is studied in Paper I. 
We choose $y = 250 \Pc$ and $\Delta y = 50 \Pc$ because the distance $250 \Pc$ is approximately 
the distance to the trailing (more distant) part of the tidal tail as seen at $45 ^\circ$ of the Pleiades 
from the  Solar System (see e.g. the upper middle panels of Fig. \ref{fevg13}).
The behaviour at different distances $y$ is qualitatively similar.
We note that during the first $\approx 70 \Myr$ of evolution, only the models containing primordial gas (CG13, CG23, and CGA) develop 
tails spanning to the distance of $\approx 250 \Pc$ of the cluster; the tails for gas-free models (CGW1 and CGW5) develop after this time.

We define the tail thickness to be an analogue 
of the Lagrangian half-number radius. 
It is the distance from the tail centre either in direction $x$ or $z$ 
which encompasses $50$~\% of the stars in the strip $|y' - y| < \Delta y$. 
The thickness of the tail in direction $x$ is shown in the left panel of Figure \ref{ftailthickness}.
The tails of models CG13 change their thickness substantially with time as the tail tilts (see also Fig. \ref{fevg13}) 
from $\approx 70 \Pc$ at maximum to $\approx 10 \Pc$ at minimum, after which it oscillates aperiodically.  
The tails of models CG23 and CGA also thicken first, reaching a maximum   $\Delta x$ at $\approx 70 \Myr$, whereupon 
they thin, and their thickness remains nearly constant after $t \gtrsim 150 \Myr$ without the 
aperiodic oscillations seen in models CG13. 
The absence of oscillations after $t \gtrsim 150 \Myr$ in models CG23 and CGA is caused by the tail II stars, 
which start dominating the tail (middle panel of \reff{fsigmaNStars}). 
The tail thickness of the gas-free models CW1 and CW5 changes smoothly with time again without 
rapid oscillations. 

The thickness of the tail in direction $z$ is shown in the right panel of Fig. \ref{ftailthickness}.
The clusters with rapid gas expulsion and $\sfe = 1/3$ show the largest amplitudes of oscillation 
(from $\approx 3 \Pc$ to $\approx 15 \Pc$ for the more massive cluster).
The models with $\sfe = 2/3$ also have  clear oscillations, but with somewhat smaller amplitudes  
(from $\approx 4 \Pc$ to $\approx 10 \Pc$ for the more massive cluster).
The models with $\sfe = 1/3$ and adiabatic gas expulsion have an even smaller difference between 
the minima and maxima (from $\approx 5 \Pc$ to $\approx 8 \Pc$ for the more massive model near the beginning), 
with decreasing amplitude with time, so the tail is almost of 
time-independent thickness $\approx 5 \Pc$ after $150 \Myr$. 
The minima of tail thickness (short black bars) occur very close to the minima predicted in Paper I, which means that 
the tail $z$ thickness oscillates with the vertical frequency $\nu$.  
The gas-free models have almost a  time-independent tail thickness of $\approx 5 \Pc$. 

The velocity dispersion in directions $x$, $y$, and $z$ in the strip $200 \Pc < y < 300 \Pc$ is shown in Figure \ref{fvelvstime}. 
The models with $\sfe = 1/3$ and rapid gas expulsion have large variations in velocity dispersion, 
with minima attained at times predicted by the semi-analytic model of Paper I for tail I.
The minima in velocity dispersions for models with $\sfe = 2/3$ are also described well by the semi-analytic model, 
but the variations between minima and maxima are substantially smaller than in the previous case. 
The models with $\sfe = 1/3$ and adiabatic gas expulsion show even 
smaller variations with almost time-independent velocity dispersions after $\approx 150 \Myr$. 
Velocity dispersions in the gas-free models increase first (up to $\approx 150 \Myr$), and then remain approximately constant 
as the number of incoming and outgoing stars is well balanced, and the structure of the tidal tail 
stays nearly time independent at given position $y$. 

The evolution of bulk velocity $v_{\rm y}$ in the $y$ direction measured in the reference frame of the cluster
at $200 \Pc < y < 300 \Pc$ is shown in the lower right panel of Fig. \ref{fvelvstime}. 
Models CG13 and CG23 evolve close to each other: $v_{\rm y}$ decreases from $\approx 8 \Kms$ at $30 \Myr$ to zero at 
$\approx 130 \Myr$, and then even becomes  negative as the tail starts approaching the cluster. 
At $\approx 170 \Myr$, $v_{\rm y}$ reaches its minimum of $\approx - 1 \Kms$, and 
then starts increasing, with positive values reached after $\approx 200 \Myr$.
The change in sign of $v_{\rm y}$ occurs along the whole tail, as  shown in fig. 7 of Paper I (left panel).
Models CGA behave qualitatively similarly with smaller absolute values of $v_{\rm y}$. 
Gas-free models CW1 behave differently: The maximum of $v_{\rm y} \approx 1.5 \Kms$ is attained near $100 \Myr$, and then it decreases 
to $\approx 0.7 \Kms$ hardly reaching negative values. 
Gas-free models CW5 have even smaller values of $v_{\rm y}$.

Although the analysis of the tail thickness and velocity dispersion evolution was instrumental in justifying the assumptions 
of the analytic model of Paper I (particularly the instantaneousness of the escape of stars forming tail I and 
that these stars escape isotropically), these quantities do not belong to the best indicators for the 
gas expulsion scenario for the Pleiades. 
Particularly, the tail thickness $\Delta x$ and $\Delta z$ varies rapidly on the timescale of the uncertainty in the Pleiades age 
(which ranges from $\approx100 \Myr$ to $\approx130 \Myr$; \citealt{Meynet1993,Basri1996,Stauffer1998}). 
For example, the tail thickness $\Delta z$ in the $z$ direction varies from $\approx 2 \Pc$ at $100 \Myr$ to $\approx 16 \Pc$ 
at $120 \Myr$ for model C10G13, which encompasses the values of $\Delta z$ for all the other models during this time. 
Moreover, in Sect. \ref{ssGMCs} we argue that the rapid oscillations in $z$ direction are likely smeared out due to 
encounters with molecular clouds. 
The argument about rapid oscillations can also be applied to the velocity dispersion $\sigma_{\rm z}$. 
In Sect. \ref{ssAccuracyGaia}, we find that velocities in the tail can be measured with an error $\gtrsim 1 \Kms$, 
so it would  hardly be possible to detect  trends in $\sigma_{\rm y}$, which reaches maxima of only $\approx 2.5 \Kms$. 
We note that contamination from field stars (see Sect. \ref{ssContamin}) contributes another uncertainty to these quantities.
This leaves us with two more promising quantities, $\sigma_{\rm x}$ and $v_{\rm y}$, where the models 
differ by several $\Kms$ for the majority of time.
However, the Pleiades are not a suitable target cluster to show differences 
in $\sigma_{\rm x}$ or $v_{\rm y}$ because both the quantities happen to be 
close to zero at the age of the Pleiades. 
These quantities might be useful for star clusters at a different age 
(for example for a cluster of the age between $150 \Myr$ and $180 \Myr$; 
we discuss a possible target cluster in Sect. \ref{ssOtherClusters}).

\subsection{Stellar mass function of the tidal tail}

\label{ssMF}

For each model, we calculate the mass function (MF) of stars at the age $t_{\rm pl}$ in the cluster 
and in the tail separately. 
The results for all stars are shown in Fig. \ref{fmfMassive}; the left panel corresponds to the lower mass clusters, 
the right panel  to the more massive clusters. 
First, we focus on the whole mass range, which is shown in the upper row of the figure.  
To contrast the MF in the cluster from that in the tail, we plot the cluster MF by solid lines, and the tail MF by dashed lines.
The MF is always normalised to one star, and it does not take into account compact objects.
For comparison, the canonical IMF \citep{Kroupa2001a} is indicated by the yellow line.
According to the adopted stellar evolutionary tracks for solar metallicity, the most massive star still on the main sequence 
at the age of $125 \Myr$ has a mass of $\approx 4.6 \Msun$, and the most massive star which has not become a compact object 
has a mass of $\approx 4.9 \Msun$, which corresponds to late B stars. 
This mass determines the upper end of the MF in the plot. 

\iffigs
\begin{figure*}
\includegraphics[width=\textwidth]{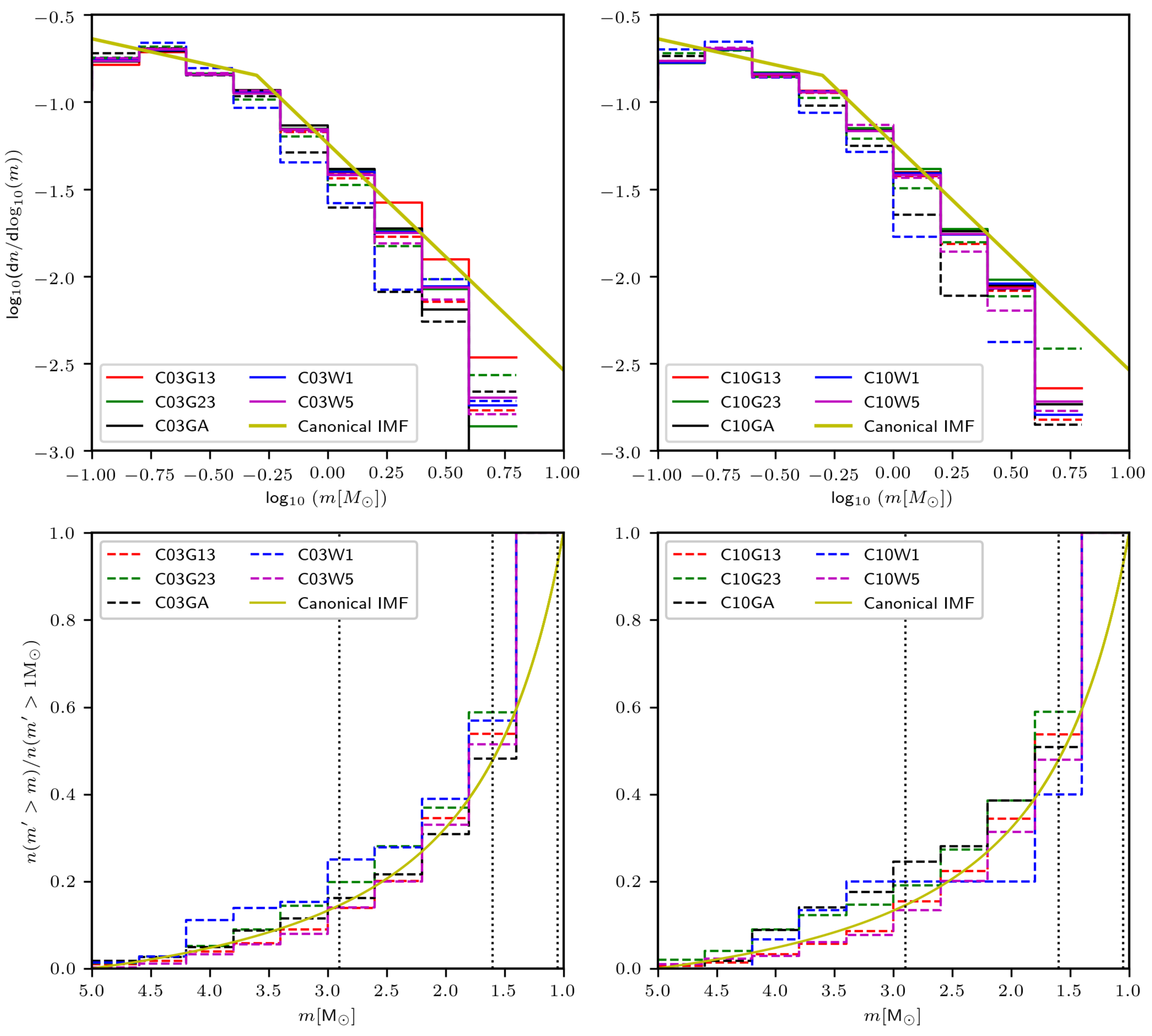}
\caption{Stellar mass function at the age of the Pleiades. 
The left column shows lower mass clusters ($M_{\rm cl}(0) \approx 1400 \Msun$);
the right column shows the more massive clusters ($M_{\rm cl}(0) \approx 4400 \Msun$).
The canonical IMF \citep{Kroupa2001a} is indicated by the yellow line.
\figpan{Upper row}: The log-log MF of all stars in the cluster (solid lines) and in the tidal 
tail (dashed lines). 
\figpan{Lower row:} The cumulative mass function for the more massive stars ($m \gtrsim 1 \Msun$) in the tidal tail. 
The masses of A0, F0, and G0 star are indicated by the vertical dotted lines.
}
\label{fmfMassive}
\end{figure*} \else \fi

\iffigs
\begin{figure*}
\includegraphics[width=\textwidth]{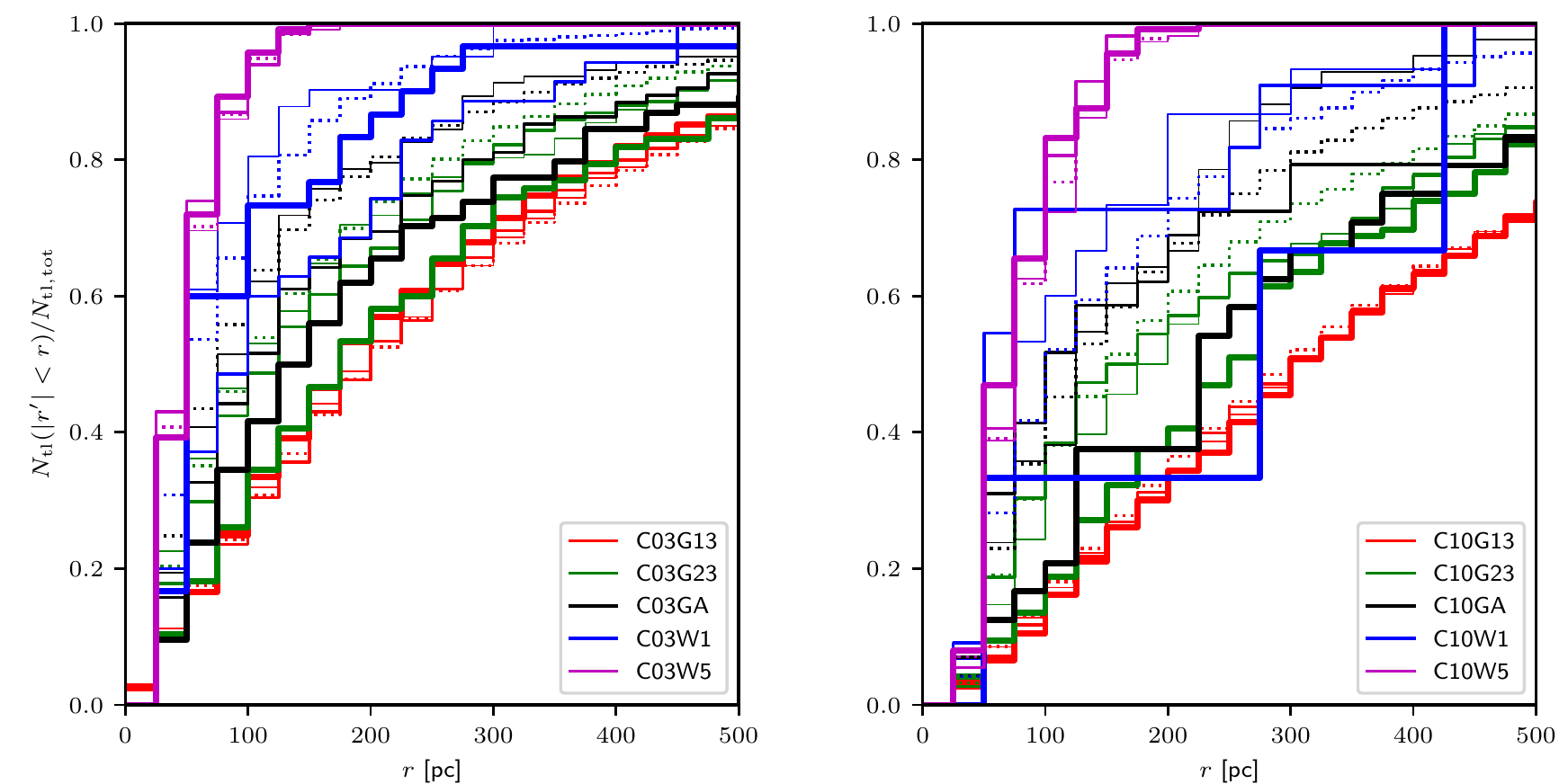}
\caption{Cumulative distribution of stellar distances $r$ from the cluster at age $t_{\rm pl}$. 
The left and right panel represents the lower mass ($M_{\rm cl}(0) \approx 1400 \Msun$) and more massive 
($M_{\rm cl}(0) \approx 4400 \Msun$) star clusters, respectively. 
We plot the histograms for the following groups of stars: A stars and earlier (solid thick lines), 
F stars (solid lines), G stars (solid thin lines), and all stars (dotted lines).}
\label{ftidalMFCumm}
\end{figure*} \else \fi

All models with fast gas expulsion have the tail MF very close to the adopted IMF. 
We also list the number fraction $f$ of the A, F, G, and K tail stars separately in Table \ref{tTailAbund}; 
we note  that the IMF of \citet{Kroupa2001a} populated from $0.08 \Msun$ to $120 \Msun$ results  in 
the number fractions of A, F, G, and K stars to be $f_{\rm A} = 0.028$, $f_{\rm F} = 0.038$, $f_{\rm G} = 0.041$, and $f_{\rm K} = 0.101$ 
\footnote{Following Table 3.13 in \citet{Binney1998}, we take the mass ranges for A, F, G, and K stars 
to be $(1.6, 2.9) \Msun$, $(1.05, 1.6) \Msun$, $(0.79, 1.05) \Msun$, and $(0.51, 0.79) \Msun$, respectively.}
.
These models also have  the number fraction of these spectral types close to the canonical values, with slightly 
underrepresented A stars ($f_{\rm A,tl} = 0.022$ instead of the canonical value $0.028$).
The agreement with the IMF is caused by the fact that a significant fraction 
of the stars forming the tail were unbound during the gas expulsion event, 
and this had occurred before mass segregation of $\approx 5 \Msun$ stars could  be established  
(see the relevant timescales in Table \ref{tsimList}), making the tail and cluster MF  very similar 
since the gas expulsion time.

\begin{table*}
\begin{tabular}{ccc|cccc|cccc|ccc}
Run name & $N_{\rm tl}$ & $N_{\rm tl,\pi/4}$ & $N_{\rm A,tl}$ & $N_{\rm F,tl}$ & $N_{\rm G,tl}$ & $N_{\rm K,tl}$ & $f_{\rm A,tl}$ & $f_{\rm F,tl}$ & $f_{\rm G,tl}$ & $f_{\rm K,tl}$ & $\frac{N_{\rm A,tl}}{N_{\rm A,cl}}$ & $\frac{N_{\rm F,tl}}{N_{\rm F,cl}}$ & $\frac{N_{\rm FGK,tl}}{N_{\rm FGK,cl}}$  \\
\hline
  C03G13 &     2437 &  971 &  53 &    91 &   101 &   217 &    0.022 &    0.037 &    0.042 &    0.090 & 2.69 & 3.85 & 4.03 \\
  C03G23 &      453 &  266 &   9 &    15 &    17 &    35 &    0.021 &    0.035 &    0.038 &    0.079 & 0.15 & 0.15 & 0.15 \\
   C03GA &      147 &   96 &   3 &     5 &     5 &    13 &    0.022 &    0.037 &    0.037 &    0.090 & 0.07 & 0.07 & 0.07 \\
   C03W1 &      109 &   79 &   2 &     3 &     3 &     7 &    0.015 &    0.026 &    0.031 &    0.068 & 0.02 & 0.02 & 0.03 \\
   C03W5 &      607 &  524 &  13 &    24 &    24 &    54 &    0.022 &    0.039 &    0.041 &    0.091 & 0.23 & 0.25 & 0.24 \\
\hline
  C10G13 &     6116 & 1712 &   133 &   234 &   272 &   550 &    0.022 &    0.038 &    0.045 &    0.091 & 1.42 & 1.52 & 1.55 \\
  C10G23 &      926 &  413 &    18 &    31 &    36 &    67 &    0.020 &    0.034 &    0.041 &    0.075 & 0.08 & 0.08 & 0.08 \\
   C10GA &      328 &  167 &     4 &     7 &    12 &    20 &    0.012 &    0.024 &    0.038 &    0.068 & 0.02 & 0.02 & 0.02 \\
   C10W1 &      173 &   94 &     0 &     3 &     4 &    11 &    0.000 &    0.018 &    0.025 &    0.074 & 0.00 & 0.01 & 0.01 \\
   C10W5 &      478 &  341 &     9 &    16 &    22 &    41 &    0.019 &    0.036 &    0.049 &    0.090 & 0.04 & 0.05 & 0.05
\end{tabular}
\caption{The number of stars of different spectral types in the tidal tail at the Pleiades age (i.e. $t = 125 \Myr$). 
From left to right, the columns list  model name;  total number of stars in the tidal tail $N_{\rm tl}$; 
 total number of tail stars $N_{\rm tl,\pi/4}$ angularly closer than $45 ^{\circ}$ to the cluster, as seen from the Solar System; 
 number $N_{\rm C,tl}$ of stars of spectral class C (where C stands for A, F, G, and K) in the tidal tail; 
 number fraction $f_{\rm C,tl}$ of stars of spectral class C in the tidal tail (e.g. $f_{\rm A,tl} = N_{\rm A,tl}/N_{\rm tl}$). 
The canonical IMF \citep{Kroupa2001a} populated from $0.08 \Msun$ to $120 \Msun$ results in
  number fractions of $f_{\rm A} = 0.028$, $f_{\rm F} = 0.038$, $f_{\rm G} = 0.041$, and $f_{\rm K} = 0.101$. 
We also compare the ratio of the tail $N_{\rm C,tl}$ to cluster $N_{\rm C,cl}$ population for stars of 
spectral types A, F, and FGK together.
The table is a result of all realisations of the models; for example, $N_{\rm A,tl} = 2$ for model C03W1 is an 
average of 13 models containing 26 A stars.}
\label{tTailAbund}
\end{table*}

Likewise, the rarefied models C03W5 and C10W5 have very long dynamical timescales with mass segregation of $\approx 5 \Msun$ 
stars taking $\approx 50 \Myr$. 
In this case, the tail MF is also expected to be close to the canonical MF, which is confirmed by our results. 
On the other hand, the models C03W1 and C10W1, which remain relatively compact, 
have a depleted tail MF by a factor of $2-3$ for stars more massive than $\approx 1 \Msun$ (see also Table \ref{tTailAbund}).
The paucity of the most massive stars in the tail is caused by the preferential evaporation of 
lower mass stars from the cluster. 
Moreover, the cluster tends to mass segregation very early (at $\approx 1 \Myr$) making it more difficult for 
more massive stars to escape (apart from high velocity escapers originating in strong interactions, but these events are rare). 

If observed, a lack of more massive stars in the tail might indicate that the cluster has not experienced gas expulsion. 
However, it appears that this cannot be easily observed for the Pleiades. 
In Sect. \ref{ssContamin} we show that at least one-half of the stars of spectral type later than F in the tidal tails 
are unlikely to be distinguished from field stars, which prevents the reconstruction of the MF below $\approx 1 \Msun$. 
This narrows our study to stars more massive than $\approx 1 \Msun$; the cumulative MF of these 
stars is shown in the lower row of Fig. \ref{fmfMassive}. 
The MF is normalised to the number of stars more massive than $1 \Msun$. 
The cumulative MFs for all models are close to the canonical MF, and the slight differences between them are 
unlikely to be detected. 
The reason why the MF of models CW1 (i.e. C03W1 and C10W1 together) is different from the other MFs in the upper panels of Fig. \ref{fmfMassive} while 
all the MFs are close to each other in the lower panels of the same figure is the normalisation adopted in the lower panels; 
the MF of models C03W1 and C10W1 is depleted by the same factor ($2-3$) regardless of mass for all stars with $m \gtrsim 1 \Msun$, 
so we cannot detect its variations from the canonical MF unless we access stars with $m < 1 \Msun$. 

Although the observed tail MF is not a good indicator for the origin of a star cluster, 
there is another indicator which appears to be more suitable for the task. 
The stars of spectral type A are the least confused by field stars, and their majority is 
likely to be detected (see the detection fraction of A stars $p_A$ in Table \ref{tContamin}). 
The census of A stars within the Pleiades is probably complete.
Table \ref{tTailAbund} shows the ratio of the A stars in the tail $N_{\rm A,tl}$ to the A stars in the cluster $N_{\rm A,cl}$. 
Gas-free models CW1 unbind very few A stars ($N_{\rm A,tl} \lesssim 0.02 N_{\rm A,cl}$). 
The number of tail A stars is substantially larger for models CW5 ($N_{\rm A,tl} \lesssim 0.23 N_{\rm A,cl}$), 
but this is still   a factor of 10 smaller than that for the gas rich models CG13.

Next, we discuss the models CW5.
Assuming that star clusters start as gas-free objects, the clusters with larger $r_{\rm h}$ evaporate faster 
forming a more populated tail
\footnote{This holds for evaporation. On the other hand, compact clusters where a handful of massive stars 
are packed to a small central volume ($\lesssim 0.1 \Pc$) eject a significant amount of 
its most massive stars \citep{Fujii2011,Oh2015,Perets2012}, 
and perhaps all O-type stars can be ejected from some clusters \citep{PflammAltenburg2006,Kroupa2018,Wang2019}. 
It appears that the more compact the cluster, the higher  the fraction of ejected early-type stars \citep{Oh2015,Wang2019}. 
However, this process does not interfere with our results for A stars because (i) ejections are related to the most massive stars, 
while A stars are more than $10 \times$ less massive than the most massive stars in our models (also note that 
runaway A stars are rare in comparison to O stars, \citealt{Gies1986,Stone1991}); (ii) if there were ejected A stars, they 
would travel at high speeds as runaways ($\gtrsim 30 \Kms$), and apart from being easily discernible in Gaia data, they 
would quickly leave (in less than $15 \Myr$) the proposed volume occupied by the tidal tail, which extends to $\lesssim 500 \Pc$.}
.
Further assuming that newborn clusters have $r_{\rm h} < 5 \Pc$ initially \citep{Marks2012,Kuhn2014,Kuhn2017}, we can set an upper limit on 
the population of A stars in the tail relative to the A star population of the cluster for gas-free clusters. 
The tail population is smaller than $0.23 \times$ that of the cluster for the less massive clusters ($M_{\rm cl}(0) = 1400 \Msun$), 
and even smaller ($0.04 \times$) for the more massive clusters ($M_{\rm cl}(0) = 4400 \Msun$). 
If tidal tails more populated than this are observed around the Pleiades, it will be 
impossible to explain their origin as a single star cluster without primordial gas expulsion or another form of disruption. 
We note that this is the upper limit on $N_{\rm A,tl}/N_{\rm A,cl}$ 
because the models with $r_{\rm h}(0) = 5 \Pc$ have $r_{\rm h}(t_{\rm pl}) \approx 5 \Pc$ (Fig. \ref{fLagrangeCl}), 
while $r_{\rm h}(t_{\rm pl}) \approx 2 \Pc$ is observed for the current Pleiades \citep{Raboud1998}. 

Next, we discuss models with gas expulsion and different SFE. 
For fast gas expulsion models CG13, the majority of A stars is located in the tail ($N_{\rm A,tl} = 2.7 N_{\rm A,cl}$ for model C03G13, 
and $N_{\rm A,tl} = 1.4 N_{\rm A,cl}$ for model C10G13). 
The tails become less numerous as the SFE increases from $1/3$ to $2/3$, 
so we extrapolate that as SFE increases from $1/3$ to $2/3$, $N_{\rm A,tl}/N_{\rm A,cl}$ decreases significantly
and $r_{\rm h}(t_{\rm pl})$ decreases slightly.
Since the tails of models CG23 have a comparable ratio $N_{\rm A,tl}/N_{\rm A,cl}$ to that of models CW5, 
the SFE around $2/3$ is the highest value of SFE that can be unambiguously distinguished from gas-free models. 
For example, a finding of $N_{\rm A,tl}/N_{\rm A,cl} = 0.5$ would imply the SFE between $1/3$ and $2/3$ and rapid gas expulsion. 
It does not appear that models CGA can be distinguished from models CG23 and CW1 by comparing the number of stars in 
the tail and in the cluster (the value of $N_{\rm A,tl}/N_{\rm A,cl}$ are comparable).  

In this comparison, we focused on the number of tail A stars because of their lower contamination by field stars. 
We note that very similar results can be obtained by counting F stars or F-K stars together 
(last two columns of Table \ref{tTailAbund}). 

Are stars of different spectral types in the tail spatially separated, or are they well mixed?
Figure \ref{ftidalMFCumm} shows the cumulative distribution of selected groups of stars as a function 
of the distance $r$ from the star cluster at the Pleiades age. 
The selected groups are A stars and earlier (thick lines), F stars (medium thick lines), G stars (thin lines);  
for reference we plot the results for all stars (dotted lines). 
Gas-free clusters as well as the clusters with $\sfe = 1/3$ and rapid gas expulsion 
(i.e. models CW1, CW5, and CG13) do not show 
any trend with stellar mass in the cumulative distribution in the tail, indicating that these 
models have well-mixed stellar populations, and the MF is independent of $r$. 
The other models (i.e. CG23 and CGA) indicate the distribution of stellar masses within the tail 
with lower mass stars concentrated closer to the cluster than more massive stars. 
This is easily explained by the structure of the tails; as we noted in Sect. \ref{sOverviewModels} (see also Table \ref{tsimList})
the tails of models CG23 and CGA are composed of stars released both due to gas expulsion and evaporation. 
Gas expulsion acts at the beginning and unbinds more and less massive stars equally; evaporation 
 preferentially unbinds lower mass stars and they travel at lower speeds, so the 
lower mass stars are located closer to the cluster. 
However, given the heavy contamination of field stars (only around 20\% to 60\% of the tail F-M stars can 
be recovered in an optimistic scenario; see Table \ref{tContamin} and Sect. \ref{ssContamin}), 
it is improbable that these subtle differences will be observed.

\section{Observational limitations}

\label{sObsLim}

We show that the confusion with field stars is a more limiting factor when observing the Pleiades tidal tail than 
the errors on Gaia measurements. 

\subsection{Accuracy of Gaia data}

\label{ssAccuracyGaia}


Which of the tidal tail morphologic or kinematic features studied above can be detected by the accuracy of measurements of Gaia 
in the case of the Pleiades star cluster?
We aim particularly at the Pleiades due to their proximity, where the Gaia data will contain 
the smallest errors; the case of more distant clusters is discussed in Sect. \ref{ssOtherClusters}.
The expected errors of Gaia measurements are calculated for the final data release
\footnote{Because we did not find the errors of transversal velocity in the literature for the final data release, 
we use the errors from the current DR2 release, which are the upper limits of these errors.}
.


The errors of the position and velocity are estimated for stars of spectral classes A0, F0, G0, and K0
at the distance of the Pleiades ($134 \pm 5 \Pc$ \citealt{Gaiac2016a}), and at the outer parts of the tail 
(which we take to be at a distance of $500 \Pc$ from the Solar System 
because the majority of tail stars is located closer than this distance for all considered 
gas expulsion scenarios, Fig. \ref{ftidalMFCumm}).
This implies a distance modulus of $5.6$ for the cluster and  $8.5$ for the outer tail. 

At the distance of the Pleiades, the apparent $G$ magnitude of an A0, F0, G0, and K0 star is 
$6.2$, $8.3$, $10.0$, and $11.2$, respectively 
(the V magnitudes of these stars were taken from Table 3.13 in \citealt{Binney1998}, and the bolometric correction 
for the Gaia $G$ magnitude from \citealt{Andrae2018}). 
In the final data release, Gaia will provide 
parallaxes with error $\sigma_{\rm \varpi} = 7 \Mas$ for stars with $3 < G < 12.09$ \citep{Gaiac2016b}, thus 
the distance to all stars in the Pleiades earlier than spectral type K2 
will be measured with an extremely high accuracy of $0.15 \Pc$, 
and Gaia will reveal even the relative distances between stars within the cluster (with half-mass radius of  $\approx 2 \Pc$). 
The error on radial velocity (the radial and transversal velocity are relative to the Sun) 
was calculated by eq. 14 in \citet{Gaiac2016b}, and it ranges from $0.5 \Kms$ for A0 stars 
to $0.7 \Kms$ for K0 stars. 
Assuming an error on the transversal velocity of $60 \Mas \; \rm{yr}^{-1}$ 
(as  for the Gaia DR2 release) and a transversal velocity of the 
Pleiades no higher than $50 \Kms$,
the error on transversal velocity is less than $0.1 \Kms$. 

For the outer parts of the tail (at a distance of $500 \Pc$), the  error on the distances 
for stars earlier than type F6, which are brighter 
than $G = 12.09$, is $1.7 \Pc$, and the errors are $2.2 \Pc$ and $4 \Pc$ for G0 and K0 stars (see eq. 4 in \citealt{Gaiac2016b}). 
The error on radial velocity is $0.7 \Kms$, $1.2 \Kms$, $1.4 \Kms$, and $4 \Kms$ for  A0, F0, G0, and K0 stars, respectively. 
The error on transversal velocity grows from $0.2 \Kms$ for A0 stars to $0.5 \Kms$ for K0 stars.  
The errors generally increase for less luminous stellar classes. 

Thus, the spatial structure of the whole tidal tail will be well resolved, and probably 
 the tail bulk velocity $v_{\rm y}$ will also be determined or constrained;  we note that 
the Pleiades appear to have the tail bulk velocity near zero at their age (see lower right panel of Fig. \ref{fvelvstime}). 
In particular, the extent of the tail is an important signpost of the gas expulsion history and can 
distinguish between different models (cf. Fig. \ref{flagrtail} and \reft{tPredTail} below).
A velocity error always larger than $0.5 \Kms$ is likely 
an obstacle in recognising the differences in velocity dispersion between the models (Fig. \ref{fvelvstime}).

\subsection{Contamination from field stars}

\label{ssContamin}

We have seen that Gaia will be able to clearly detect the morphological signatures of the tidal tail. 
However, there are field stars occupying the same area in the phase space as the tail stars contaminating the tail.  
Now we discuss whether the tail is distinct enough to be detected above the background of field stars. 
We consider two sources of contaminating stars: a smooth Galactic population represented by a Schwarzschild 
distribution function and a more concentrated Hyades-Pleiades (HyPl) supercluster, which is a prominent 
stream located close in the velocity space to the Pleiades. 
These calculations present only an order of magnitude estimate as the stellar distribution 
in velocity space shows significant substructures, which are absent in the smooth analytic models. 

First, we discuss the Galactic population. 
We adopt the properties of the Galactic field stars from the current iteration of the 
Besan\c{c}on model (\citealt{Czekaj2014}, see also \citealt{Robin2003}). 
This model divides the stellar population of the Galactic disc in seven groups according to their 
age, spanning the whole lifetime of the thin disc, which is assumed to be $10 \Gyr$. 
Each group has its mass density and velocity dispersion. 
For an order of magnitude estimate we neglect the contribution from the halo and the thick disc to the 
stellar density near the Galactic midplane because the contribution is less than $10$ \%. 
We take into account only the stellar population of the thin disc. 
To obtain an upper estimate on the contamination, we consider  model A of \citet{Czekaj2014} 
because it yields higher background stellar density. 
We also convert the provided mass density, $\rho_{\rm field}$, to the stellar number density, $n_{\rm field}$, by assuming 
the present-day mass function to be populated only in the interval $(0.08 \Msun, 1.0 \Msun)$; this is approximately 
the interval populated by the oldest stellar group due to stellar evolution.  
This together with the IMF of \citet{Kroupa2001a} results in $3.5$ stars per one $\Msun$ of the population. 
We note that we adopt a conservative estimate because this choice increases the number density of contaminating stars. 

Each age group $j$ of the Besan\c{c}on model has its density $\rho_{\rm field,j}$ (in units of $\Msun \Pc^{-3}$), thus the number density 
of field stars of spectral class C in this group 
is $n_{\rm C,field,j} = 3.5 f_{\rm C} \rho_{\rm field,j}/ \Msun$, 
where $f_{\rm C}$ is the number fraction of stars of spectral class C in a given stellar population. 
The velocity distribution of the group is modelled with a Schwarzschild distribution function, $f_{\rm schw}$ (see eq. 4.156 in 
\citealt{Binney2008}), where we assume that the ratio of radial to vertical velocity dispersion 
  is $\sigma_{\rm z} /\sigma_R = 0.5$ \citep{Binney2008}, and that the ratio is independent of the Galactocentric radius and the age group. 
We note that the distribution function $f_{\rm schw} = f_{\rm schw}(\sigma_{\rm U}, \sigma_{\rm V}, \sigma_{\rm W})$ 
depends on the velocity dispersions, and thus that it is different for each age group
\footnote{Since the purpose of this section is to provide an order of magnitude estimate, we neglect vertex deviation.}
.

How many field stars happen to occupy the same volume in velocity space as the Pleiades? 
For this we adopt a Galactocentric velocity vector of the Pleiades star cluster of $(U_{\rm pl}, V_{\rm pl}, W_{\rm pl}) = (-4.7, -23.8, -8.9) \Kms$ \citep{Chereul1999}. 
The velocity dispersion, $\sigma_{\rm tl}$, of the Pleiades tail is up 
to $\approx 2 \Kms$ for the majority of our models (Fig. \ref{fvelvstime}), 
so if we include a $2 \sigma_{\rm tl}$ margin of the tail velocity dispersion (where the vast majority of the tail stars lie), we 
obtain a velocity limit on stars associated with the Pleiades tail to $v_{\rm tl} = 4 \Kms$. 
For comparison, we also calculate contamination within $3 \sigma_{\rm tl}$ ($v_{\rm tl} = 6 \Kms$), which is 
higher, and we list the value in brackets.
Accordingly, we estimate the fraction of the stars of group $j$, which contaminate the Pleiades, as 
\begin{equation}
f_{\rm cont,j} = \frac{\int\limits_{\substack{||\tilde{\mathbf{V}}|| < v_{\rm tl}}}
\! \! \! \! \! \! \! \! f_{\rm schw}(U,V,W,\sigma_{U,j}, 
\sigma_{V,j}, \sigma_{W,j}) \dd U \dd V \dd W}{\! \! \! \! \! \! \! \! \int\limits_{(-\infty, \infty)^3} \! \! \! \! \! \! \! \! f_{\rm schw}(U,V,W,\sigma_{U,j}, \sigma_{V,j}, \sigma_{W,j}) \dd U \dd V \dd W},
\label{eFracSchw}
\end{equation}
where
\begin{equation}
||\tilde{\mathbf{V}}|| = \left\{ (U-U_{\rm pl})^2 + (V-V_{\rm pl})^2 + (W-W_{\rm pl})^2 \right\}^{1/2}. 
\label{edelV}
\end{equation}
The number density of contaminating stars of spectral type C from all the $N_{\rm G}$ groups is then
\begin{equation}
n_{\rm C,cont} = \Sigma_{j=1}^{N_{\rm G}} n_{\rm C,field,j} f_{\rm cont,j}.
\label{eContamin}
\end{equation}

The critical density for contamination occurs when 
\begin{equation}
n_{\rm C,cont} \approx n_{\rm C,tl},
\label{eCritDensContamin}
\end{equation}
where $n_{\rm C,tl}$ is the number density of stars of spectral class C in the tidal tail.
Here we assume that spectral types in the tail are distributed according to the adopted MF  
and are well mixed (i.e. $f_{\rm C}$, the number fraction of stars of spectral type C in the population, is independent 
of the position within the tail). 
This assumption is justified by the results presented in  Sect. \ref{ssMF}.
Accordingly, we adopt $n_{\rm C,tl} = n_{\rm tl} f_{\rm C} $. 
Equations (\ref{eContamin}) and (\ref{eCritDensContamin}) then imply the condition for threshold contamination 
\begin{equation}
\Sigma_{j=1}^{N_{\rm G,C}} n_{\rm C,field,j} f_{\rm cont,j}/f_{\rm C} \approx n_{\rm tl},
\label{eCritDensContaminEq}
\end{equation}
where the upper sum limit, $N_{\rm G,C}$, denotes the oldest age group containing main sequence stars of spectral type C. 

Naturally, we focus on the brightest stars.
However, the most massive stars are absent due to stellar evolution, so for the Pleiades,
spectral classes A, F, G, and K are the most interesting.
For simplicity we assume that the younger groups 1-4 (age up to $3 \Gyr$) 
in the model of \citet[][see their table 4]{Czekaj2014} contain 
stars of spectral class A0-K9, while the older groups (5-7) in the model contain 
stars of spectral class F0-K9. 
This gives $n_{\rm tl} \approx 7\times 10^{-8} \; \Pc^{-3} (5 \times 10^{-7} \; \Pc^{-3})$ for A stars, 
and $n_{\rm tl} \approx 1 \times 10^{-5} \; \Pc^{-3} (4 \times 10^{-5} \; \Pc^{-3})$ for F, G, and K stars for 
$v_{\rm tl} = 4 \Kms (6 \Kms)$. 

It should be noted that the meaning of $n_{\rm tl}$ is   the number density of all stars in the tail where the density of the selected spectral class (A or later than A, 
which is well mixed in the tail) equals the 
density of the background population (either for the Galaxy or for the HyPl stream) of the selected spectral class. 
As a rough estimate, the areas in the tidal tail as traced by a given stellar type can be recognised above background 
when the density of all stars in the tail lies above $n_{\rm tl}$. 
There is a lower threshold value for A stars than FGK stars  not only because   their presence is restricted to 
the younger age groups, but also because the younger groups have a smaller velocity dispersion, and thus 
are more concentrated towards the origin of the velocity space $(U, V, W)$. 
The threshold value of $n_{\rm tl}$ for FGK stars, which is more constraining than that for A stars, 
is indicated by the white contours in Figs. \ref{fevg13} through \ref{fevWo5}. 
The area enclosed by these contours shows the part of the tail with acceptable contamination from field 
stars according to the Besan\c{c}on model. 
We also note that the contamination for spectral types F, G, and K (and also M) are the same regardless of the luminosity because 
we assume these spectral types are represented in each of the seven Besan\c{c}on age groups (F stars lifetimes are shorter than the 
age of the Universe, so their contamination is a bit lower than in the later spectral types, but we neglect this 
in this first study).

Table \ref{tContamin} lists the expected `observed' half-number radius $r_{\rm h}$ of the tail, as calculated from 
A stars and earlier, and for FGK stars located above the threshold density $n_{\rm tl}$. 
The table also lists the fraction $p$ of tail stars located 
above $n_{\rm tl}$ (done for A stars and earlier and from FGK stars separately). 
The left section of the table contains the results of the analysis for all tail stars (i.e. in the case of no contamination), 
while the middle section contains results for the contamination of the Besan\c{c}on model. 
For comparison, we also list the total number of A stars and earlier ($N_{\rm A+}$) and FGK stars ($N_{\rm FGK}$) within the tail. 


The fraction $p_{A+}$ of A stars and earlier above the contamination threshold is very high (more than 80\% for all models having more 
than a handful of A stars, i.e. for all models where their analysis makes sense from a statistical point of view). 
The vast majority of A stars is not significantly contaminated; they could be identified, and the kinematics of tails in these models
could be reconstructed very well. 
Thus, their half-number radius, $r_{\rm h,A+}$, is comparable to the uncontaminated cluster (cf. columns 7 and 3). Thus,
A stars  offer a reliable tool for distinguishing the different initial conditions for the Pleiades. 

On the other hand, FGK stars are significantly contaminated, and only 40\% to 70\% could be identified. 
Although the heavy contamination might skew the sample so that important kinematic signposts are biased, 
they  will still allow us to distinguish at least between models with an $\sfe = 1/3$ and the other models 
by the number of stars detected in the tail.

\begin{table*}
\begin{tabular}{ccccc|cccc|cccc}
Run name & $N_{\rm A+}$ & $r_{\rm h,A+}$ & $N_{\rm FGK}$ & $r_{\rm h,FGK}$ & $p_{A+}$ & $r_{\rm h,A+}$ & $p_{\rm FGK}$ & $r_{\rm h,FGK}$ & $p_{A+}$ & $r_{\rm h,A+}$ & $p_{\rm FGK}$ & $r_{\rm h,FGK}$ \\
 & & [pc] & & [pc] &  & [pc] &  & [pc] &  & [pc] &  & [pc]  \\
\hline
  C03G13 &       68 &   198 &      192 &   217 & 1.00 &   198 & 0.73 &   165 & 0.60 &   149 & 0.51 &   129  \\
  C03G23 &       13 &   232 &       32 &   122 & 0.84 &   163 & 0.56 &    70 & 0.33 &   103 & 0.44 &    59  \\
   C03GA &        4 &   169 &       10 &    75 & 0.84 &    98 & 0.50 &    46 & 0.23 &    37 & 0.30 &    36  \\
   C03W1 &        2 &   131 &        5 &    76 & 0.95 &    86 & 0.60 &    52 & 0.40 &    20 & 0.40 &    47  \\
   C03W5 &       16 &    56 &       48 &    57 & 1.00 &    56 & 0.98 &    57 & 1.00 &    56 & 0.98 &    56  \\
\hline
  C10G13 &      171 &   363 &      505 &   350 & 1.00 &   362 & 0.63 &   248 & 0.46 &   211 & 0.37 &   178  \\
  C10G23 &       26 &   289 &       67 &   192 & 0.88 &   248 & 0.48 &    98 & 0.25 &   119 & 0.30 &    78  \\
   C10GA &        6 &   380 &       18 &   154 & 0.80 &   287 & 0.39 &    69 & 0.07 &    15 & 0.22 &    59  \\
   C10W1 &        0 &     0 &        6 &   106 &    - &     0 & 0.50 &    66 &    - &     0 & 0.17 &    58  \\
   C10W5 &       11 &    90 &       38 &    87 & 1.00 &    90 & 0.97 &    86 & 0.95 &    88 & 0.92 &    84
\end{tabular}
\caption{Tidal tail contamination due to field stars at the age of the Pleiades. 
To provide an idea of the statistics, we also list the total number of A and earlier stars ($N_{\rm A+}$) and 
the total number of F, G, and K stars ($N_{\rm FGK}$) in the tail per one simulation. 
We calculate the half-number radius for each of the group separately ($r_{\rm h,A+}$ and $r_{\rm h,FGK}$).
The values in Cols. 2-5 are for the tail in total, i.e. without any contamination taken into account. 
Columns 6-9 list the fraction $p_A+$ of stars earlier than A, and $p_{\rm FGK}$ for F, G, and K stars together which 
are located above the contamination threshold for the Besan\c{c}on model, which is $n_{\rm tl} \approx 7 \times 10^{-8} \Pc^{-3}$ 
for A stars, and $n_{\rm tl} \approx 1 \times 10^{-5} \Pc^{-3}$ for FGK stars. 
The half-number radius for both subgroups calculated for stars that are above the density threshold 
is denoted $r_{\rm h,A+}$ and $r_{\rm h,FGK}$. 
Columns 10-13 are the same as Cols. 6-9, but for the case of the higher contamination, i.e. HyPl stream 
and the Besan\c{c}on model together. 
In this case, the contamination threshold is $n_{\rm tl} \approx 2 \times 10^{-5} \Pc^{-3}$ for A stars and earlier, 
and $n_{\rm tl} \approx 3 \times 10^{-5} \Pc^{-3}$ for F, G, and K stars. 
The contamination threshold for both the Besan\c{c}on model and HyPl stream is taken for velocity $|v_{\rm tl}| \lesssim 4 \Kms$ 
around the $(U_{\rm pl}, V_{\rm pl}, W_{\rm pl}) = (-4.7, -23.8, -8.9) \Kms$ velocity of the Pleiades. 
The results are averaged over 4 simulations for the more massive clusters, and over 13 
simulations for the lower mass clusters.}
\label{tContamin}
\end{table*}

In the Galactic model of \citet{Czekaj2014}, it is assumed that the stellar velocity distribution 
is smooth and represented by the Schwarzschild DF. 
However, there are distinct stellar streams, which cause overdensities in the UVW space. 
The streams typically encompass many star clusters, OB associations and stars \citep{Eggen1958,Chereul1998,Chereul1999,Famaey2005}. 
In particular, the Pleiades star cluster is a part of the Hyades-Pleiades (HyPl) stream. 
Although we demonstrated that the smooth stellar background of a Schwarzschild DF does not 
obscure the Pleiades tail, it is also necessary to estimate the magnitude of the background star contamination 
due to the HyPl stream. 

The HyPl stream is approximated by a 3D  Gaussian centred at $(U,V,W) = (-30.3, -20.3, -4.8) \Kms$ with 
a velocity dispersion of $(\sigma_{\rm U}, \sigma_{\rm V}, \sigma_{\rm W}) = (11.8, 5.1, 8.8) \Kms$ \citep[][their table 2]{Famaey2005}.
As for the estimate for the Besan\c{c}on model, we assume $v_{\rm tl} = 4 \Kms$, but also list the contamination threshold
for $v_{\rm tl} = 6 \Kms$ in  brackets. 
To obtain an order of magnitude estimate, we assume that all spectral types follow the same Gaussian distribution, 
and that any stellar spectral type has the same relative abundance in the stream as in the Galaxy's field population. 
With 392 giants identified within the stream out of a total of 6030, and a number stellar density of 
the total Galactic population of $n = 0.14 \Pc^{-3}$ (\citealt[][]{Czekaj2014}, their table 7), it implies that 
the number density of the HyPl stream is $n = 0.009 \Pc^{-3}$, and the number density of the HyPl stream satisfying 
the velocity criterion is $n_{\rm tl} = 2.0 \times 10^{-5} \Pc^{-3} (7.0 \times 10^{-5} \Pc^{-3})$. 

The threshold $n_{\rm tl}$ for the HyPl stream is  a factor of 2 higher than for the Besan\c{c}on model for FGK stars. 
In contrast to the Besan\c{c}on model, the density threshold for A stars is the same as that for FGK stars  because of the assumption that 
the HyPl stream is too young for stellar evolution to  significantly deplete its population of A stars \citep{Chereul1998}.

While the contamination due to the Besan\c{c}on model is  a lower estimate for the real contamination, 
we attempt to put an upper estimate by adding the contamination due to the Besan\c{c}on model and HyPl stream together.
The results for an `observation' of the combined contamination are shown in the right section of Table \ref{tContamin}. 
The detection fraction for A stars is significantly lower (around $30\%-60$\% for most of the models), 
and also the `observed' $r_{\rm h,tl}$ radii smaller than the real ones (the most affected models CGA and CGW1 
have observed $r_{\rm h,tl}$ radii smaller by a factor of several). 
This implies that it will be unlikely to distinguish between the gas-free models and models with adiabatic gas expulsion 
from the structure of the tidal tail; however, models with rapid gas expulsion and $\sfe = 1/3$ (and also $\sfe = 2/3$) 
show conspicuously longer tails. 
We note that for the combined contamination 
there is almost no difference between the tails in A and FGK stars.

The combined contamination threshold is shown in Figs. \ref{fevg13} through \ref{fevWo5} 
(actually three times the combined value to get an upper estimate) by the black contours. 
Even for the upper estimate, the parts of the tail located above this threshold 
density show clear differences between models CG13, CG23, and the gas-free model CW1. 

\section{Predictions for the Pleiades tidal tail}

\label{sPredictions}

\iffigs
\begin{figure*}
\includegraphics[width=\textwidth]{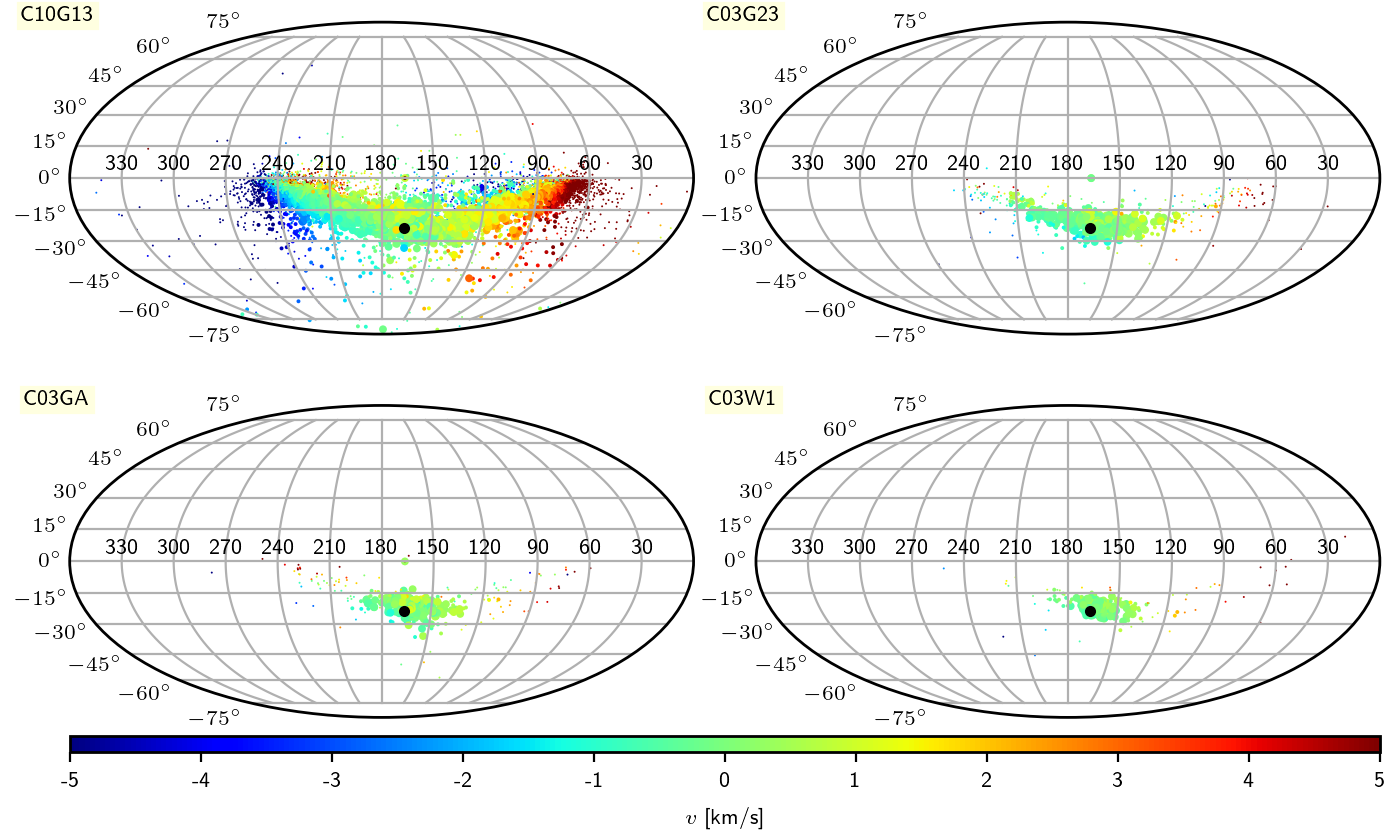}
\caption{Position of the tidal tail in Galactic coordinates as seen from the Solar System. 
The shown models correspond to those with the closest match for the Pleiades star cluster at the present time.
The position of the Pleiades at $(l, b) = (166.57, -23.5)$ is indicated by the black circle. 
The colour represents the radial component of the tail velocity relative to the cluster as seen from the Solar System 
(blue for  an approaching star). 
To obtain the velocity relative to the Solar System, this velocity should be added to the relative 
velocity between the Pleiades and the Solar System. 
The symbol size shows the stellar number density in the tail at the star; the smallest points are 
for $n_{\rm tl} \lesssim 2.0 \times 10^{-5} \Pc^{-3}$.}
\label{fProjSky}
\end{figure*} \else \fi

To apply the results of the previous sections to the Pleiades star cluster, we find the 
values of the initial parameters (mass and radius) which evolve at age $t_{\rm pl}$ to a cluster with similar mass, 
half-mass radius, and velocity dispersion as the current Pleiades. 
We attempted to set the initial conditions so that models with a different scenario of 
gas expulsion (e.g. $\sfe = 1/3$ or gas-free clusters) evolve to the state resembling the current Pleiades. 
Although the ensuing clusters are similar, the tidal tails often demonstrate remarkable differences 
for different gas expulsion scenarios, opening the possibility to distinguish between 
the gas expulsion event in the case of the Pleiades after its tidal tail is identified in Gaia data.

For the comparison we assume the current mass of the Pleiades cluster to be $ = 740 \Msun$ \citep{Converse2008,Pinfield1998}, 
the half-mass radius $r_{\rm h} = 2 \Pc$ \citep{Limber1962,Raboud1998}, and the 3D velocity dispersion 
$\sigma_{\rm cl} =0.5 \Kms$ \citep{Jones1970,Raboud1998}. 
The three quantities are compared with those for each of the models with different initial conditions, which are plotted in 
Figs. \ref{fLagrangeCl} and \ref{fsigmaNStars} (lower panel). 

Both models CG13 have the expected $r_{\rm h}$; the agreement is also close for $\sigma_{\rm cl}$ (with model 
C03G13 being very close, and C10G13 approximately two times above $\sigma_{\rm cl}$). 
The lower and more massive models have $M_{\rm cl}(t_{\rm pl})$ 
lower and higher ($280 \Msun$ and $1800 \Msun$, respectively) than that of the Pleiades. 
This indicates that the model with $\sfe = 1/3$ and rapid gas expulsion and initial mass somewhere between the initial 
masses for these models (i.e. $1400 \Msun$ and $4400 \Msun$) evolves very close to the Pleiades. 
Accordingly, the predictions for the tail morphology based on the assumption of $\sfe = 1/3$ and rapid gas expulsion 
lies somewhere between the two models, C03G13 and C10G13. 
As we saw in Sect. \ref{sTail}, the tails of these models are close to each other, so this prediction is robust. 
Following similar reasoning, models C03G23 and C03GA also evolve to the state close to the current Pleiades, so these models 
also give a prediction for the tail morphology, but for different assumptions about the gas expulsion scenario.

The radius for the gas-free models CW1 is close to the observed value, but even the lower mass model C03W1 
results in a cluster a factor of $\approx 2$ more massive than the Pleiades. 
This suggest that if the Pleiades started as a gas-free cluster, its mass was approximately $2 \times$ smaller 
than that of model C03W1, and the associated tidal tail is now less numerous and a bit shorter than in this model. 
The decrease in mass in comparison to model C03W1 also lowers $\sigma_{\rm cl}$, bringing it closer to the observed value. 
The half-mass radius of models CW5 is too large to be compared with the Pleiades; on the other hand, 
the compact gas-free models CW02 expand to $r_{\rm h} \approx 2 \Pc$ by $t_{\rm pl}$. 
Extrapolating from this, if the Pleiades formed 
as an initially gas-free cluster, its initial half-mass radius could hardly be larger than $1 \Pc$, 
but it could be substantially smaller. 
The results of the most interesting observable quantities of the tail are summarised in Table \ref{tPredTail}.

\begin{table*}
\begin{tabular}{ccccc}
cluster model & $r_{\rm h,tl} [\Pc]$ & $N_{\rm A+,tl,obs}$ & $N_{\rm A+,tl,obs}/N_{\rm A+,cl}$  &  MF tail  \\
\hline
$\sfe = 1/3$ fast GE & $150 \Pc - 350 \Pc$ & $40 - 170$ & $0.7 - 2.7$  & canonical \\
$\sfe = 2/3$ fast GE & $100 \Pc - 200 \Pc$ & $4 - 11$ & $0.05 - 0.15$ & canonical \\
$\sfe = 1/3$ slow GE & $40 \Pc - 100 \Pc$ & $1 - 4$ & $0.02 - 0.07$ &  canonical \\
$\sfe = 1$           & $<20 \Pc - <90 \Pc$ & $<1 - <2$ & $ < 0.02$ & depleted in stars of $m \gtrsim 1 \Msun$
\end{tabular}
\caption{Summary of the predictions of different gas expulsion scenarios for the most robust observational quantities. 
The quantities are  tidal tail half-number radius, $r_{\rm h,tl}$, as traced by A stars; 
 number of A stars and earlier in the tidal tail, $N_{\rm A+,tl,obs}$, located above 
the contamination thresholds ($N_{\rm A+,tl,obs} = p_A N_{\rm A+,tl}$) and thus observable; 
 ratio of the observable A stars and earlier in the tail to those  in the cluster, $N_{\rm A+,tl,obs}/N_{\rm A+,cl}$; 
and   shape of the present-day tail MF. 
The lower (upper) bounds were obtained by taking the minima (maxima) of corresponding quantities in Table \ref{tContamin} 
for the upper (lower) estimate 
of the contamination threshold for A stars. 
The ratio $N_{\rm A+,tl,obs}/N_{\rm A+,cl}$ is based on $N_{\rm A,tl}/N_{\rm A,cl}$ in Table \ref{tTailAbund} assuming 
that all A stars and earlier in the cluster are observed.} 
\label{tPredTail}
\end{table*}

The relative position of stars in the tail as seen from the Solar System is shown in Fig. \ref{fProjSky}. 
Model C10G13 forms a very rich tail, which stretches to large distances from the cluster, with 
some stars located even at the Pleiades anticentre. 
The number of stars $N_{\rm tl,\pi/4}$ angularly closer to the Pleiades than $45 ^{\circ}$ (Table \ref{tTailAbund}) 
is only $\approx 30$\% of the total tail population. 
The asymmetry of the tail (the leading tail is angularly longer than the trailing one) arises because the leading tail is closer 
to the Sun. 
Models C03G23 and C03GA form substantially shorter and less numerous tails, with the majority ($\approx 60$ \%) of stars 
located closer than $45 ^{\circ}$ to the cluster. 
Gas-free model C03W1 forms the shortest and poorest tail. 
Thus, to test the gas expulsion scenario, 
it is insufficient to search for tail stars only in the vicinity of the Pleiades,
but the search should extend practically all over the sky.

\section{Discussion}

\label{sDiscussion}

\subsection{Dependence on other parameters}

As in numerical modelling, we could explore only a limited region of the parameter space. 
In this study, we focus on changing the SFE and the timescale, $\tau_{\rm M}$, of gas removal. 
One parameter, which we do not change, is the duration of the embedded phase $t_{\rm d}$, 
which is related to the dynamics of the ultra-compact HII regions, and which we  always take as $0.6 \Myr$. 
The range of values of this parameter has not been studied extensively
(the only study known to us where $t_{\rm d}$ is varied is the work of \citealt{Banerjee2013}, but they studied a massive cluster, 
$ \approx 10^{5} \Msun$, with a median two-body relaxation time $t_{\rm rlx}$ much longer than $t_{\rm d}$). 
Observationally, the possible range of $t_{\rm d}$ is estimated to be
from almost zero up to several million years \citep{Lada2003,Ballesteros2007}. 
This time is comparable to the mass segregation timescale $t_{\rm ms}$ for $3 \Msun$ stars, 
which is $\approx 3 \Myr$ (see Table \ref{tsimList}).
Clusters undergoing longer embedded phase mass segregate, with more pronounced mass segregation 
in lower mass clusters because of their shorter relaxation timescale.


Mass segregation (either dynamically induced or primordial) might have a strong impact on the MF of the tail, where 
the more massive stars (including those of spectral type A) are located close to the cluster centre at the time of 
gas expulsion, and the majority of them might be retained, skewing the tail MF to lower masses, and 
reducing substantially the number of tail A stars and possibly also the extent, $r_{\rm h,tl}$, measured in A stars. 
However, the extent of the tail in stars later than A is likely large and comparable to the extent of the tail 
studied in the present (non-segregated) models.
An exploration of these parameters might be an objective of a future work.

\subsection{Role of molecular clouds}

\label{ssGMCs}


To model the gaseous potential, we resort to the same approximation that is usually adopted 
when dealing with a gas component in the code: the gas distribution is spherically symmetric, 
and the centre of the gas sphere coincides with the centre of the cluster. 
This is perhaps an acceptable approximation for the infrared dark cloud which forms the cluster, 
but the rest of the surrounding giant molecular cloud (GMC) is not modelled. 
However, the total mass of the GMC exceeds the cluster mass at least by a factor of $100$. 
As the cluster emerges and destroys its parental infrared dark cloud, it continues dissolving the whole GMC 
(other star clusters formed within the GMC do the same). 
During the process, the newborn cluster is exposed to the changing gravitational field of the GMC. 

To do an order of magnitude estimate, consider an idealised model consisting of a point mass GMC of mass $10^5 \Msun$ located 
at a distance of $20 \Pc$ from a cluster of mass $4.4 \times 10^3 \Msun$. 
The resulting tidal radius of the cluster in the field of the cloud is around $5 \Pc$, which is 
substantially less than the tidal radius imposed by the Galaxy, which is $\approx 24 \Pc$. 
Although the external GMC field acts only for several Myr after the cluster formed, 
it is likely that it has a large impact on cluster dissolution. 
However, the only conceivable method for addressing this issue is 
a radiative magnetohydrodynamical simulation including realistic modelling of the cluster dynamics, 
a task which has been gradually approached (e.g. \citealt{Girichidis2011,Bate2014,Gavagnin2017}) 
but  not  satisfactorily solved, and which falls beyond the scope of the present work.

We demonstrated in Paper I that the minima in oscillations of tail I are well described by the analytic model. 
Here we confirm that the analytic model can also be used  for the minima of the tails for 
the scenarios with rapid gas expulsion and $\sfe \lesssim 2/3$.  
In the analytic model the tail thickness  
oscillates aperiodically in the direction $x$ and periodically in the direction $z$ 
with minima attained at time events, which are solely functions of Galactic frequencies $\omega$, $\kappa$, and $\nu$. 
However, these models are based on a smooth model for the gravitational potential of the Galaxy. 
The potential of the real Galaxy is substantially more substructured, mainly due to GMCs. 
As tail stars accelerate in the gravitational field of a GMC, their trajectories get curved, which causes shifts of the phases 
(i.e. $\alpha$ and $\zeta$ in eqs. 2 of Paper I) of the affected stars. 
The spread of phases smears the deep minima of tail thickness $\Delta x$ and $\Delta z$ as seen in Fig. \ref{ftailthickness}, 
possibly limiting the usefulness of $\Delta x$ and $\Delta z$ as indicators for the gas expulsion scenario.

\subsection{Implications of first Gaia results}

\label{ssGaiaResults}

Although no observation has been published about the Pleiades tidal tail yet, 
we are aware of three open star clusters, where extended tidal tails have already been observed: 
the \object{Hyades} \citep{Roser2019a,Meingast2019}, the \object{Praesepe} \citep{Roser2019b}, and \object{Melotte~111} \citep{Furnkranz2019,Tang2019}, 
which are of age $\approx 650 \Myr$, $\approx 800 \Myr$ and $\approx 700 \Myr$, respectively. 
All the observations reveal S-shaped tidal tails similar to tail II without a noticeable tail I.  
Can these tails already give us  a clue about the SFE experienced by these clusters? 

The longest simulations run only up to $600 \Myr$, which is less than the age of these clusters, 
so we can provide only an order of magnitude estimate. 
Progenitors to the observed clusters were also probably less massive than the available models. 
We study the tail properties at $y = 150 \Pc$ because it is the maximum distance for which the observed tails were detected. 
From Fig. 6 in Paper I it follows that while the density of tail II increases with time, 
the density of tail I decreases, and
the populations of both tails equilibrate near $y = 150 \Pc$ at $t \gtrsim 500 \Myr$.  
The figure shows the number density projected onto line $y$. 
Because tail I is typically broader in directions $x$ and $z$, its volume density 
is   a factor of several lower than that of tail II, and tail I would be more contaminated and more difficult 
to detect at that time. 
Thus, the available observations, which feature the S-shaped tail II are not likely to show 
tail I even if it were present. 
This suggests that younger clusters, where tail I is denser, are more suitable candidates for probing the conditions 
during gas expulsion. 

In this context, it is interesting to mention the work of \citet{Meingast2019}, who detect 
three stars comoving with the Hyades, but located outside the expected body of its S-shaped tail 
(upper left panel of their fig. 1). 
Although these stars could belong to tail I (apart from the possibility that they are contaminating sources),
we defer further speculations before observations of tails around younger star clusters become available.

\iffigs
\begin{figure*}
\includegraphics[width=\textwidth]{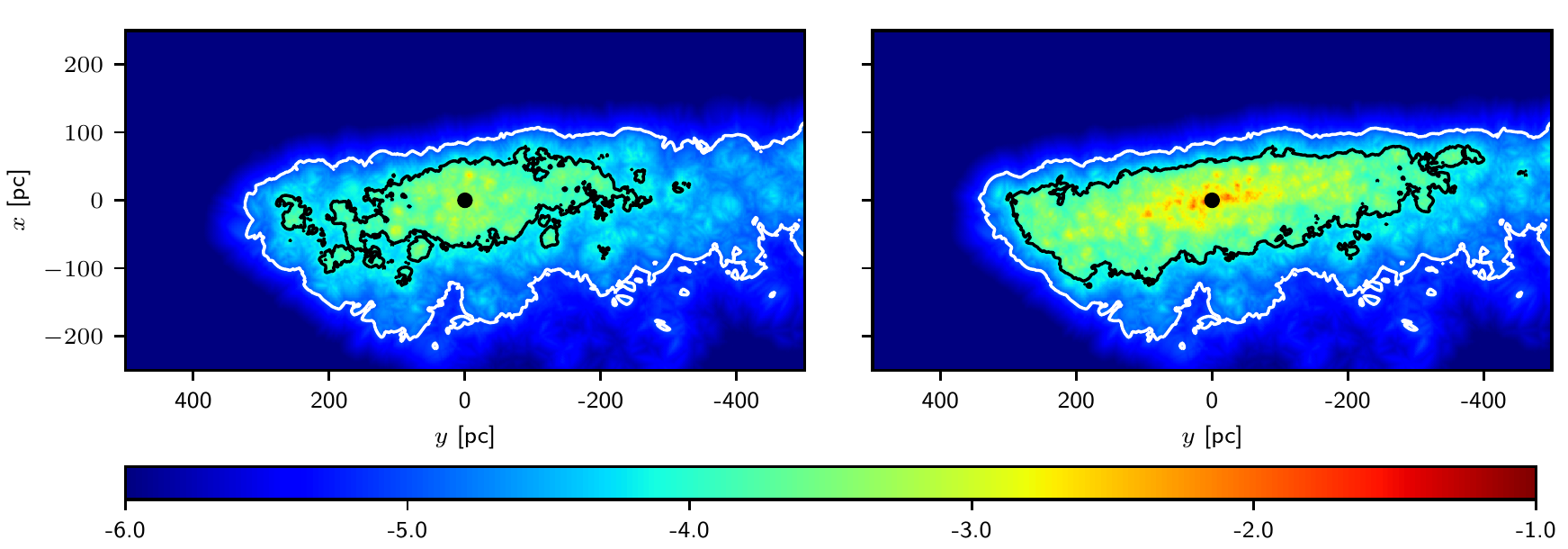}
\caption{
The stellar surface density for the star-forming molecular cloud and the most massive star cluster it formed for 
a cluster assuming $\sfe = 100$\% (left panel), and $\sfe = 33$\% (right panel) at the age of $125 \Myr$.
We plot only the stars whose velocities differ less than by $3 \Kms$ from the star cluster. 
In the vicinity of the cluster ($r \lesssim 100 \Pc$), which is represented by the black dot, the stars formed within a cloud are almost 
uniformly distributed while the stars released in the tidal tail are elongated approximately along the $y$-axis. 
The colourscale and contours have the same meaning as in \reff{fevg13}. 
}
\label{fsfreg_sf}
\end{figure*} \else \fi

\subsection{Application to other star clusters}

\label{ssOtherClusters}

Some properties of tail I might be manifested better on clusters of a different age than the Pleiades. 
During the early evolution of star clusters, models with rapid gas expulsion 
and low $\sfe \approx 33\%$ first rapidly expand, and then recollapse. 
At the maximum of their expansion, clusters reach a half-mass radius of $5$ to $10 \Pc$ (Fig. \ref{fLagrangeCl}), 
which means a low density for a young age ($5-20 \Myr$). 
Observations of young expanded clusters, and also the state during the recollapse, when the stellar velocity vectors  
converge back to the cluster centre, which clearly distinguishes them from expanding OB associations, 
would be a piece of evidence for low SFEs. 

Another property which seems unique to models with rapid gas expulsion is the motion of tail I 
between $\approx 130 \Myr$ and $180 \Myr$ back towards the cluster (lower right panel of Fig. \ref{fvelvstime}). 
In contrast, tail II always expands outwards from the cluster \citep{Kupper2008}. 
The appropriate target would be a cluster of age $\approx 160 \Myr$ and of a higher mass (so that the tail 
velocity is large). 
A possible cluster fulfilling these criteria is \object{Stock~2} with an age of $\approx 170 \Myr$, distance $\approx 320 \Pc$ 
and current mass $\approx 6000 \Msun$ \citep{Piskunov2008}. 
Assuming that the cluster formed with an $\sfe \approx 33\%$ and rapid gas expulsion, the semi-analytic model (eq. 20 in Paper I)
calibrated to model C10G13 predicts a maximum velocity $v_{\rm y}(r_{\rm h,tl}) \approx -3 \Kms$ during the tail contraction. 
In contrast, if Stock~2 formed as a gas-free cluster with $r_{h}(0) = 1 \Pc$, it produces only tail II 
stars which form a tail with a positive expansion velocity (lower right panel of \reff{fvelvstime}). 
We note that if the age of Stock~2 is determined correctly, the difference in $v_{\rm y}$ should be 
detected as the expected error of the final Gaia release in radial and transversal 
velocity for A stars is $0.7 \Kms$ and $0.2 \Kms$, respectively.

\subsection{Contamination by the stars formed in the same star-forming cloud}

It is likely that during its formation, the cluster was a part of an extended star-forming region. 
The star-forming region formed other stars and clusters, which have partially or completely disintegrated by the present time. 
These stars were subjected to the same Galactic tidal forces as the cluster in question, and they are located nearby in the position-velocity space, 
forming an extended envelope around the cluster. 
The envelope might be confused with tidal tail I even if the cluster does not form the tail. 
Is it possible to distinguish tail I from this envelope?

To perform an order of magnitude estimate, we adopt the following model. 
Consider a spherical cloud of radius $50 \Pc$, mass $10^5 \Msun$ and velocity dispersion of $6 \Kms$, 
which is a representative GMC in the Galaxy \citep{Heyer2009}. 
The cloud transforms 5\% of its total mass to stars \citep{Evans2009}, with the most massive cluster having initially $M_{\rm cl}(0) = 1200 \Msun$. 
We assume that the most massive cluster survives to the present time, while all the other star formation occurs 
in isolation at uniform density throughout the cloud. 
The latter assumption is to impose the worst-case scenario because forming the other stars in clusters would facilitate their separation in the 
position-velocity space from tail I of the cluster in question. 
We consider two simplified models for the cluster. The first  assumes that the cluster forms with an SFE of 100\%, so it does not form 
tail I, and we neglect tail II as well because it would contain only up to $\approx 100$ stars in total (see \reff{fsigmaNStars}). 
The other model assumes that the cluster forms with a lower SFE of 33\% and rapid gas expulsion, which forms a rich tidal tail. 
In both models the star cluster is located at rest at the centre of the GMC to place it at the point of the highest contamination. 
The trajectories of the stars formed throughout the cloud, as well as the ones released from the cluster, are calculated by eqs. 2 of Paper I.

Figure \ref{fsfreg_sf} compares the stellar surface density of the two models at time $t_{\rm pl}$. 
We take into account only stars with relative velocity to the cluster smaller than $3 \Kms$, which corresponds to the initial 
velocity dispersion of the cluster. 
The tidal tail I, which is formed in the model with $\sfe = 1/3$ (right panel), can be clearly 
recognised in the vicinity of the cluster ($r \lesssim 100 \Pc$) above the more uniform density distribution 
resulting from the star formation throughout the GMC, which is plotted in the left panel. 
The tail is also located well above the density thresholds for field stars, which were introduced in \refs{ssContamin}. 
Thus, it is likely that if young star clusters form tidal tail I, 
it can be separated from the more distributed star formation occurring in the same cloud.

\section{Summary}

\label{sSummary}

We attempt to shed some more light on the question of the star formation efficiency (SFE) in young star clusters by 
simulating these systems with the collisional code \nbdvid.
We vary the SFE in the range from 33 \% to 100 \%, and also experiment with the rapid (impulsive) and slower (adiabatic) 
expulsion of the natal gas out of the cluster. 
Since we aim to provide predictions particularly for the Pleiades cluster due to its proximity, 
the mass of the models is chosen so that the initial mass of the Pleiades likely fits between the 
mass of the stellar component of the modelled clusters. 
Each of the simulations is run for a cluster mass of $1400 \Msun$ and $4400 \Msun$, 
and the results are an average of 13 realisations of the lower mass cluster and 4 realisations of the higher mass cluster. 
The star clusters are integrated in the external field of the Galaxy, 
with a particular emphasis on the formation and evolution of its tidal tail. 
The tidal tail forms self-consistently from stars escaping the cluster due to gas expulsion, dynamical evaporation, and 
dynamical ejections, and we integrate the tail stars together with the cluster for $300 \Myr$.

The evolved tails (at the Pleiades age of $125 \Myr$) differ significantly for models with different 
scenarios for gas expulsion.
Models with rapid gas expulsion and a low SFE of $1/3$ form rich and extended tails, the number of stars in the tail 
decreases substantially as the SFE increases to $2/3$, with models with an SFE of $100$ \% having the least populous tails. 
Models with adiabatic gas expulsion and an SFE of $1/3$ form significantly poorer and shorter tails than the 
models with rapid gas expulsion. 
The morphology of the tail during the first $200 \Myr$ of evolution released from clusters 
of different gas expulsion scenarios 
is shown in Figs. \ref{fevg13} through \ref{fevWo5}. 
While the tidal tail for models with rapid gas expulsion and an SFE of $1/3$ is dominated by stars from tail I, 
the tidal tail for the models with adiabatic gas expulsion and an SFE of $1/3$ consists of tails I and II with 
fairly balanced populations. 

The present-day mass function of the Pleiades tail is close to the canonical MF for all the models except the ones 
with 100\% SFE and $r_{\rm h}(0) = 1\Pc$, the MF of which is depleted for stars of $m \gtrsim 1 \Msun$. 
The cluster with $\sfe = 1/3$ and rapid gas expulsion has an MF that is well mixed within the tail, 
so the relative abundance of more massive stars is independent on the distance from the cluster. 
On the other hand, the clusters with $\sfe = 2/3$ and rapid gas expulsion, and with $\sfe = 1/3$ and adiabatic 
gas expulsion show slight variations of the MF along the tail with less massive stars, which are released due to gradual 
dynamical evaporation, are more abundant closer to the cluster (\reff{ftidalMFCumm}).
However, given the contamination from the field stars, it is unlikely that these subtle 
differences will be detected. 
We note that the simulations feature clusters with no primordial mass segregation; primordially mass segregated 
clusters likely have tail I depleted from more massive stars. 

We estimate the severity of observational limitations in detecting the kinematic signatures which distinguish 
between different models. 
The distance errors on Gaia measurements are so small that all stars earlier than at least class K0 will be 
placed accurately within the whole tail. 
The velocity errors are more restrictive, increasing with distance from the Solar System and for later spectral classes 
(the typical error on tail velocity is $\approx 1.5 \Kms$), 
making tail velocity structures more difficult to observe in general. 
Moreover, the bulk velocity, $v_{\rm y}$, of the tail stars along the tail and the radial velocity 
dispersion, $\sigma_{\rm x}$, are nearly zero at the current age of the Pleiades. 

We also provide an order of magnitude estimate of the contamination due to the field stars. 
We consider two models for field star contamination: the Schwarzschild velocity distribution with the values of the Besan\c{c}on model, 
and the Hyades-Pleiades stream. 
The contamination is far more restrictive than the Gaia accuracy as only a fraction of the tail stars are 
located above the density threshold for the contamination, and these stars are located predominantly 
closer to the cluster, underestimating the true extent of the tidal tail. 
Contamination from the field stars likely smears out the subtle differences between gas-free models 
and models with adiabatic gas expulsion (even though they have a relatively low $\sfe = 1/3$); 
however, even the upper estimate on contamination does not prevent us from distinguishing 
between the 
models with rapid gas expulsion with $\sfe = 1/3$ from these of rapid gas expulsion 
with $\sfe = 2/3$ and these from the gas-free models. 

Taking the observational limitations into account, the most promising indicators 
for distinguishing the initial conditions for the Pleiades at their current age are as follows 
(see Table \ref{tPredTail} for details):


\begin{itemize}
\item
The ratio of the number of stars in the tail to the number of stars in the cluster.  
The prominent tails of the models with $\sfe = 1/3$ and rapid gas expulsion contain $0.7$ to $2.7$ times more 
A stars than the cluster.
Increasing the SFE to $2/3$ decreases the ratio to $0.05-0.15$. 
Clusters with adiabatic gas expulsion have poorer tails with this ratio being $0.02-0.07$.
Gas-free clusters form the poorest tails with this ratio being smaller than $0.02$. 
These ratios are similar for stars of other spectral types, but A stars are likely to be the least contaminated 
by the field population.
We do not expect the differences between the models with $\sfe \gtrsim 2/3$ and the models 
with adiabatic gas expulsion to be detected due to contamination. 
On the other hand, if clusters experience early rapid gas expulsion with $\sfe \lesssim 2/3$, they are 
accompanied by rich detectable tails.
\item
The half-number radius $r_{\rm h,tl}$ of the stars which can be observed in the tail: 
As the SFE increases from 1/3 to 2/3 and $100$\%, the observable value of $r_{\rm h,tl}$ decreases from $150 - 350 \Pc$ 
to $100 - 200 \Pc$ and $< 20 \Pc - < 90 \Pc$ for models with rapid gas expulsion. 
Models with adiabatic gas expulsion have an observable $r_{\rm h,tl} \approx 40 \Pc - 100 \Pc$.
\item
Models with rapid gas expulsion (for $\sfe \approx 1/3$) produce  
two kinematically distinct tails: tail II is short, poor, and S-shaped 
enveloped 
within a large, thick, and numerous tail I. 
Only the models with rapid gas expulsion form large enveloping tails. 
The enveloping tails have many stars in quadrants where the product $xy$ is positive; 
these quadrants are almost devoid of stars for the gas-free models (middle row of \reff{fevg13} and Fig. 4 
in Paper I).
\end{itemize}

The early rapid gas expulsion with a relatively low SFE ($\approx 30$ \%) leaves other kinematic 
signatures, which can be observed for clusters of different ages than the Pleiades. 
Particularly, at a young age ($\approx 5 - 20 \Myr$), the clusters first expand and then recollapse. 
The stellar velocities converge towards the cluster centre during recollapse, which would distinguish them 
from the expansion or disorganised motions as often seen in OB associations. 
The expansion of the tail is non-monotonic where periods of expansion are interspersed by periods 
of contraction (the first contraction occurs between $\approx 130 \Myr$ and $\approx 180 \Myr$). 
In contrast, the tails from evaporating stars are always expanding \citep{Kupper2008}. 
An observation of a star cluster surrounded by a large tail moving towards the cluster would indicate 
a low SFE and rapid gas expulsion.

\begin{acknowledgements}

The authors thank to Sverre Aarseth for developing and maintaining the N-body family of codes as well 
as for making them freely available and well documented. 
The authors are indebted to Sverre Aarseth also for his useful comments on the manuscript. 
FD would like to thank Ladislav \v{S}ubr and Bruno Jungwiert for sparkling his interest in N-body dynamics, 
to Sambaran Banerjee for his help with settings of \nbdvid, and to Stefanie Walch for her encouragement to work on the project 
while being employed within her group.
FD acknowledges the support by the Collaborative Research Center 956 ("The conditions and impact of star formation"), 
sub-project C5.
We thank an anonymous referee for the constructive report.

\end{acknowledgements}

%
%

\bibliographystyle{aa} 
\bibliography{clusterTail} 

\listofobjects


\begin{appendix}
\section{Detail of half-mass radius evolution}

Figure \ref{flagrcldetail} zooms-in on the half-mass cluster radius evolution during the first $10 \Myr$ after cluster formation.

\iffigs
\begin{figure*}[h!]
\includegraphics[width=\textwidth]{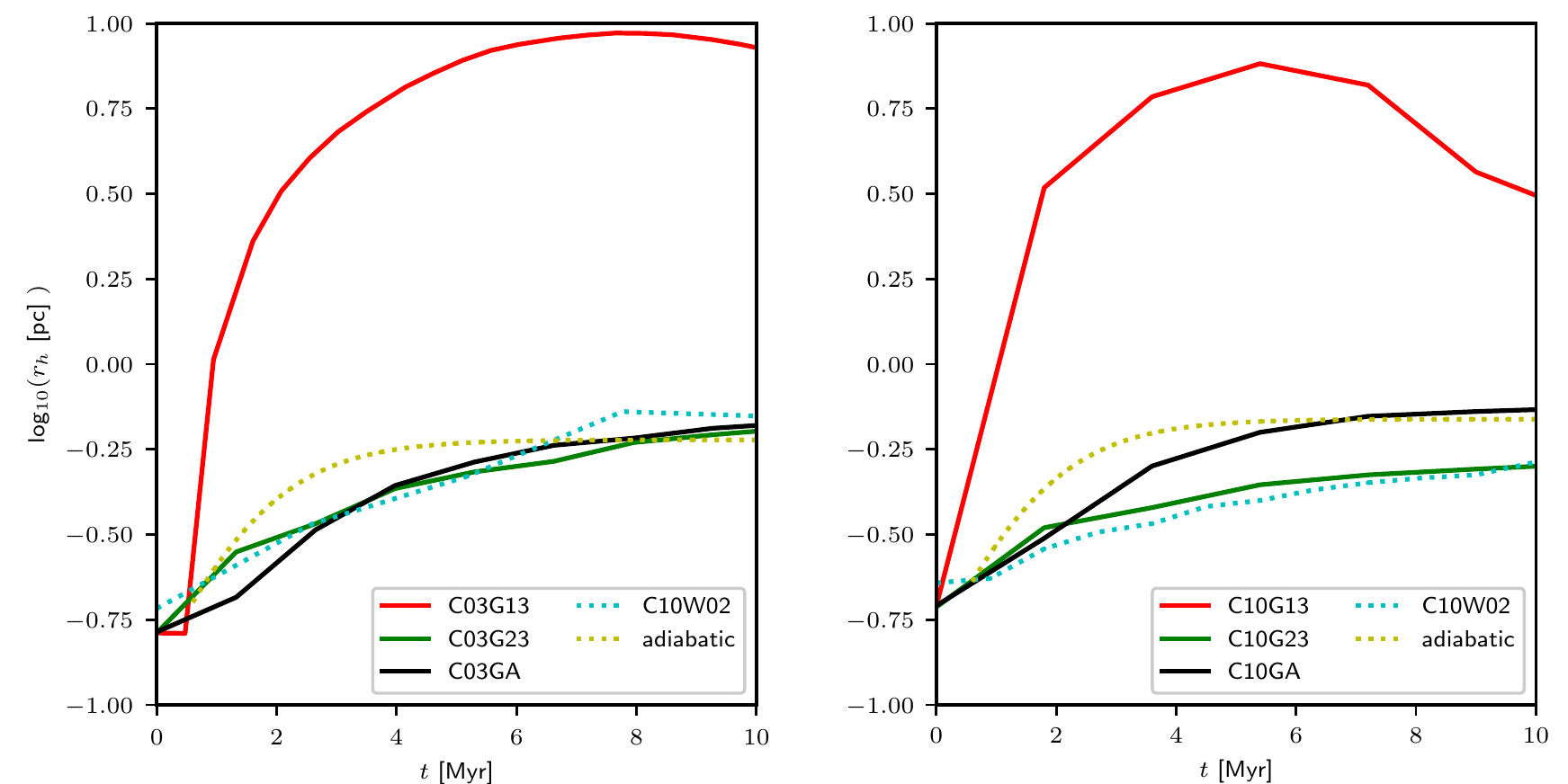}
\caption{Details of the evolution of the star cluster half-mass radius $r_{\rm h}$ for selected models. 
The results for lower mass clusters ($M_{\rm cl}(0) = 1400 \Msun$) are shown in the left panel, 
  for more massive clusters ($M_{\rm cl}(0) = 4400 \Msun$)   in the right panel. 
An estimated half-mass radius for a system assuming conservation of adiabatic invariants, 
i.e. $r_{\rm h}(0) \approx r_{\rm h}(0) (M_{\rm cl}(0) + M_{\rm gas} (0))/(M_{\rm cl}(0) + M_{\rm gas} (t)) \approx 3 r_{\rm h}(0)/(1 + 2 \exp{(-(t - t_{\rm d})/\tau_{\rm M})})$, 
where the second equality follows from eq. 27 in Paper I and $\sfe = 1/3$, 
with the gas expulsion timescale $\tau_{\rm M} = 1 \Myr$, is indicated by the yellow dotted line.}
\label{flagrcldetail}
\end{figure*} \else \fi

\end{appendix}

\end{document}